# Design and performance of the Focusing DIRC detector

B. Dey[a,1], M. Borsato[b], N. Arnaud[b], D.W.G.S. Leith[c], K. Nishimura[c], D.A. Roberts[d],
B.N. Ratcliff[c], G. Varner[e], J. Va'vra[c,*]

[a]*University of California, Riverside, CA 92521, U.S.A.*

[b]*Laboratoire de l'Accélérateur Linéaire, Centre Scientifique d'Orsay, F-91898 Orsay Cedex, France*

[c]*SLAC National Accelerator Laboratory, Stanford, California 94309 USA*

[d]*University of Maryland, College Park, MD 20742, U.S.A.*

[e]*University of Hawaii, Honolulu, HI 96822, U.S.A.*

## Abstract

We present the final results from a novel Cherenkov imaging detector called the Focusing DIRC (FDIRC). This detector was designed as a full-scale prototype of the particle identification system for the *SuperB* experiment [1], and comprises 1/12 of the *SuperB* barrel azimuthal coverage, with partial photodetector and electronics implementation. The prototype was tested in the SLAC Cosmic Ray Telescope which provided 3-D tracking of cosmic muons with an angular resolution of ~1.5 mrad, a position resolution of 4-5 mm, a start time resolution of 70 ps, and muon tracks above ~2 GeV tagged using an iron range stack. The fused silica focusing photon camera was coupled to a full-size BaBar DIRC bar box and was read out, over part of the full coverage, by 12 Hamamatsu H8500 multi-anode photomultipliers (MaPMTs) providing 768 pixels. We used waveform digitizing electronics to read out the MaPMTs. We give a detailed description of our data analysis methods and point out limitations on the present performance. We present results that demonstrate some basic performance characteristics of this design, including: (a) single photon Cherenkov angle resolutions with and without chromatic corrections, (b) signal-to-noise (S/N) ratio between the Cherenkov peak and background, which primarily consists of ambiguities of the possible photon paths from emission along the track to a given pixel, (c) dTOP = TOP$_{measured}$ - TOP$_{expected}$ resolutions (with TOP being the photon Time-of-Propagation in fused silica), and (d) performance of the detector in the presence of high-rate backgrounds.

## 1. Introduction

The DIRC [2] detector at the BaBar experiment provided excellent π/K particle identification performance [3]. Based on this success, we have been pursuing an R&D program to develop a compact and fast detector for future particle identification systems, especially those that operate at very high rates [4,5]. One such concept, the Focusing DIRC (FDIRC) [6-8], is a 3D device capable of measuring not only 2D coordinates of each Cherenkov photon with an angular resolution similar to the BaBar DIRC, but of also measuring each photon's Time-of-Propagation (TOP) along the fused silica bar with 100-200 ps single-photoelectron timing resolution. Though this is substantially better timing resolution than BaBar DIRC (with a timing resolution of ~1.6 ns/photon), it is less stringent than that of the TOP counter [9] in Belle-II (with a resolution of ~100 ps/photon) or the TORCH counter [10] at LHCb (which is planning a resolution of ~50 ps/photon). The FDIRC utilizes a full 3D readout, with resolutions in each dimension that enhance the PID performance. It can determine the Cherenkov angle with very good precision independent of the achieved timing performance, which improves the robustness of the design in practical experimental environments. This requirement determines the size of the focusing optics and the number and sizes of the pixels required. The improved timing resolution and large number of pixels compared with BaBar DIRC also provides better background rejection, an important feature for next generation detectors operating in high-intensity environments, for example at a *Super B-Factory*, while the improved timing could also modestly enhance the particle separation in some regions of phase space through particle time-of-flight (TOF). Precise timing would also allow one to correct for the chromatic dispersion contribution to the Cherenkov angle resolution and thus improve the single-photon angle measurement substantially by 1-2 mrad compared to uncorrected distributions ($\sigma_{Chromatic}$ ~4-4.5 mrad - see chapter 7.4). By using photon detectors with small pixel size, one can reduce the size of the expansion volume by up to a factor of 10 relative to the BaBar DIRC, while maintaining a similar spatial resolution (for comparison, the BaBar DIRC used a very large expansion volume of ~6000 liters because the size of each detector element was ~2.5 cm). The smaller geometric size together with better timing resolution would improve background suppression by nearly two orders of magnitude.

---

[1] Currently at Physik Institut, University of Zurich.
[*] Corresponding author: Tel.: +1-650-926-2658; fax: +1-650-926-4178; e-mail: jjv@slac.stanford.edu.

The first FDIRC prototype, employing a single DIRC fused silica bar and a cylindrical mirror placed in a mineral oil expansion volume, was constructed and operated in a test beam in 2005, 2006, and 2007 [6-8], and later tested in the SLAC cosmic ray telescope (CRT) [11] using 3D tracks [12]. It was the first RICH detector to successfully correct for the chromatic error using timing. In this article, we describe final tests of a new FDIRC design [13,14], employing fused silica optics and a full size BaBar DIRC bar box. This prototype was designed as the final demonstration test for the barrel DIRC detector of the *SuperB* experiment [1] that was to be built in Frascati.

The new design had to utilize unmodified BaBar DIRC bar boxes [3] due to resource and cost constraints. This imposed a number of restrictions on possible focusing optics designs, which prevented achieving the best possible Cherenkov angle resolution. Nevertheless, the optics and photon detector choices of the present FDIRC prototype deliver similar Cherenkov angle resolutions as the BaBar DIRC in a significantly reduced camera volume. Performance could be further improved if the pixel size were reduced to 3 × 12 mm$^2$.

Other experimental applications that might benefit from this work are the GLUEX DIRC [15] and the LHCb TORCH [10] detectors, as they both intend to re-use the BaBar DIRC bar boxes. GLUEX plans to use very similar optics to the FDIRC design, presented in this paper, to measure both the photon Cherenkov angle and time. The TORCH detector will use new optics and concentrate on measuring particle TOF very precisely (~10 ps/track), but achieving this timing resolution will require an appropriately accurate measurement of the Cherenkov angle in order to correct for chromatic effects. We believe that this work may also be useful to the PANDA barrel DIRC [16], and the future barrel DIRC planned for the electron-ion collider (EIC) [17].

## 2. Description of the FDIRC prototype

The particle identification system for the *SuperB* experiment, like other high-intensity Super Flavor Factory experiments, needed to cope with much higher luminosity-related background rates, by about two orders of magnitude, compared to previous generation experiments such as BaBar, while maintaining similar physics performance. The basic design strategy for dealing with these increased rates was to make the photon camera smaller and faster, as shown in Figs. 1a-e. The design decision was to use a new photon camera-imaging concept based on cylindrical mirror focusing optics [13]. The focusing element was machined from radiation-hard Corning 7980 fused silica. In addition to the reduced background sensitivity gained from the geometric reduction in size, the use of fused silica also reduces background sensitivity, especially to neutrons, compared to the water-filled expansion volume used in the BaBar DIRC. The improved timing not only helps with background suppression, but can also be used to measure and correct for the chromatic dispersion, thus improving performance. By including a cylindrical mirror that focuses light onto the photodetector plane, we can remove the pinhole-size (i.e. bar thickness) component of the angular resolution in one dimension of the fused silica bar that was one of the limiting factors in the BaBar DIRC design.

The FDIRC prototype was constructed utilizing a spare bar box from the BaBar DIRC [3]. This is a complete bar box consisting of 12 fused silica bar radiators surrounded by a light and gas tight enclosure and support structure. The individual fused silica bars have nominal dimensions of 35 mm width × 17.25 mm height × 4900 mm length, as shown in Figs. 1f and 1g. They each are attached to an individual wedge of approximately the same width as the bar, and all 12 bar assemblies are coupled to the bar box window. Figures 1d and 1e show the new optical components: a new wedge and a large focusing block (FBLOCK) [14]. The dimensions of the FBLOCK were nominally 560 mm height × 217 mm length × 422 mm width. The new wedge was coupled to the bar box window with optical epoxy (50~75 μm thick Epotek 301-2). The FBLOCK was then coupled to the new wedge with RTV (1mm thick Shin-Etsu 403). The entire assembly was housed in a light-tight mechanical support structure that is connected to the bar box, as shown in Fig. 1c.

(a)

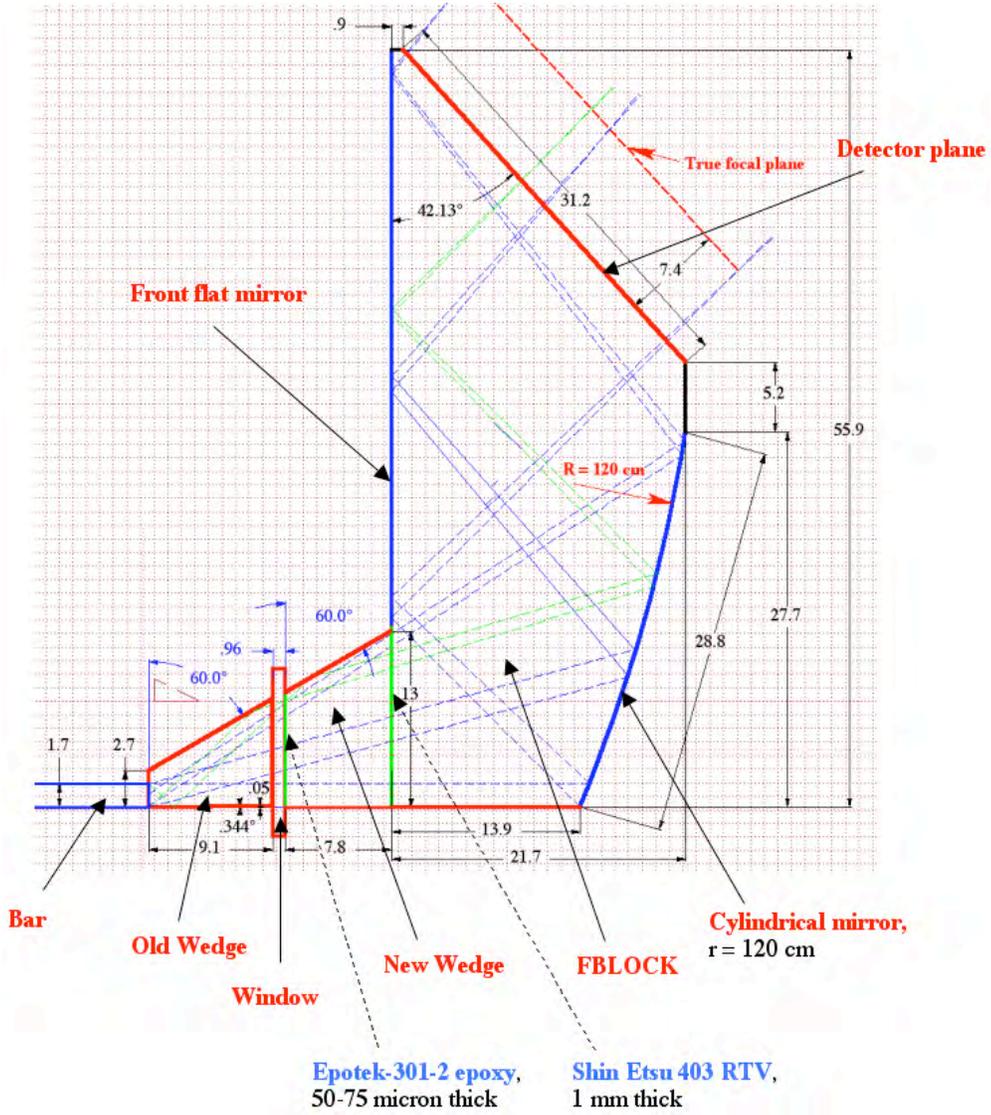

(b)

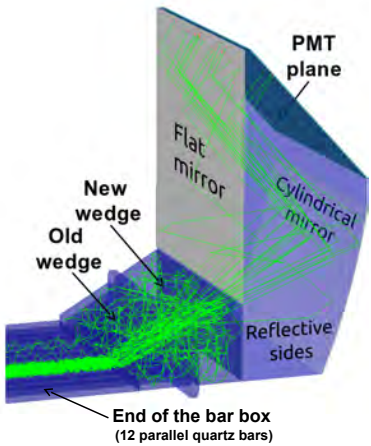

(c)

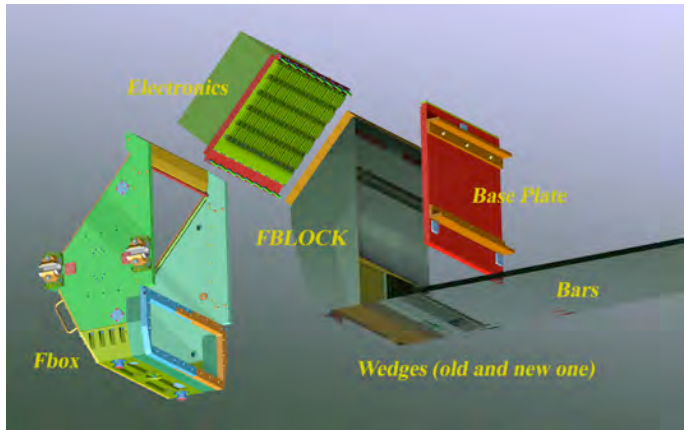

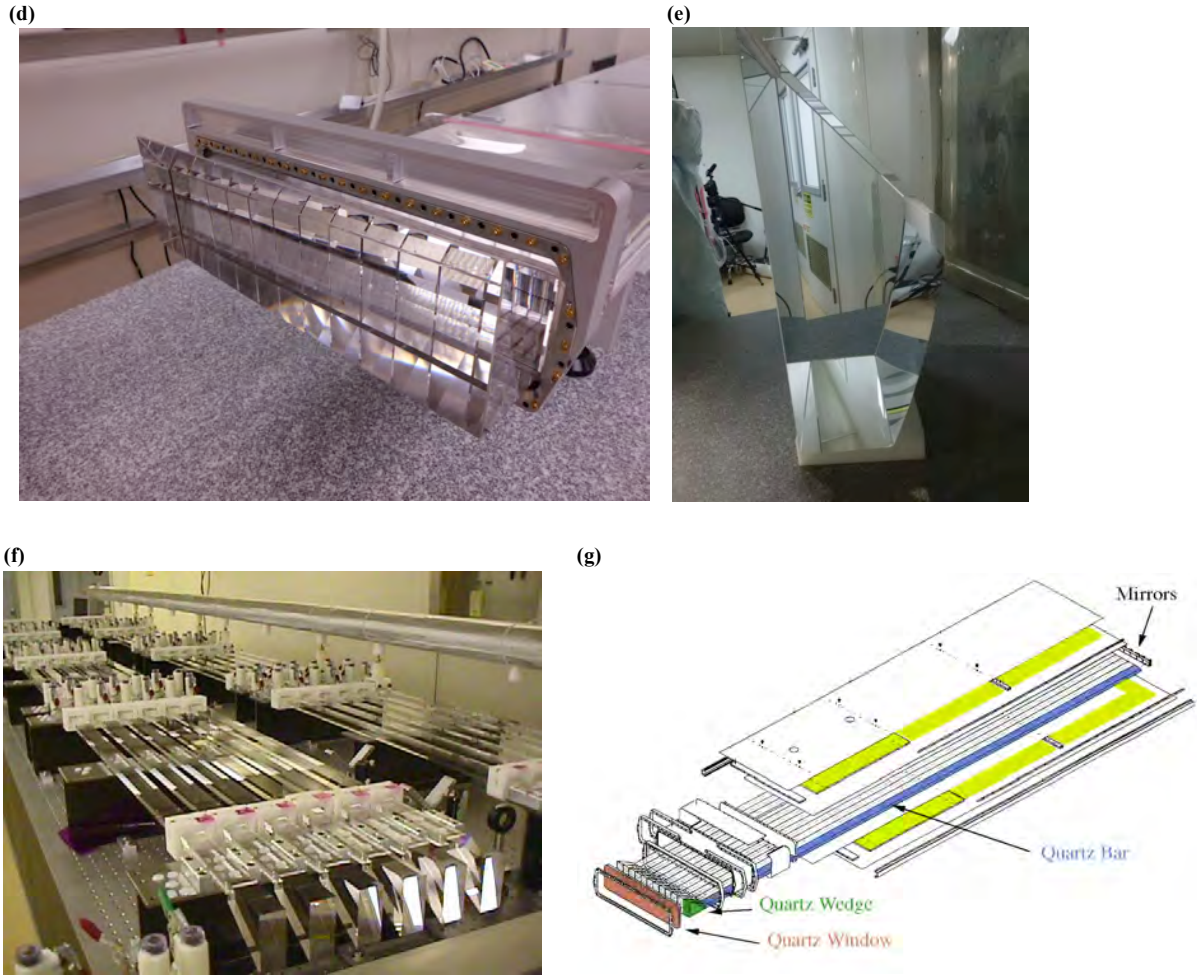

**Fig. 1** The FDIRC prototype: (a) Side view of the focusing optics showing major ray traces and dimensions in cm. (b) A 3D view of the new optics showing a GEANT4 simulation of photon tracks (there is 1 old wedge/bar, 12 bars/bar box, and one full-width new wedge/bar box). (c) Assembly drawing of various parts of the FDIRC photon camera. (d) A new wedge glued to the bar box window. (e) FBLOCK of this prototype. (f) DIRC Fused silica radiator bars ready to be moved into the bar box (this picture was taken during the BaBar DIRC box construction). (g) Assembly drawing of the BaBar DIRC bar box, which is used in this prototype.

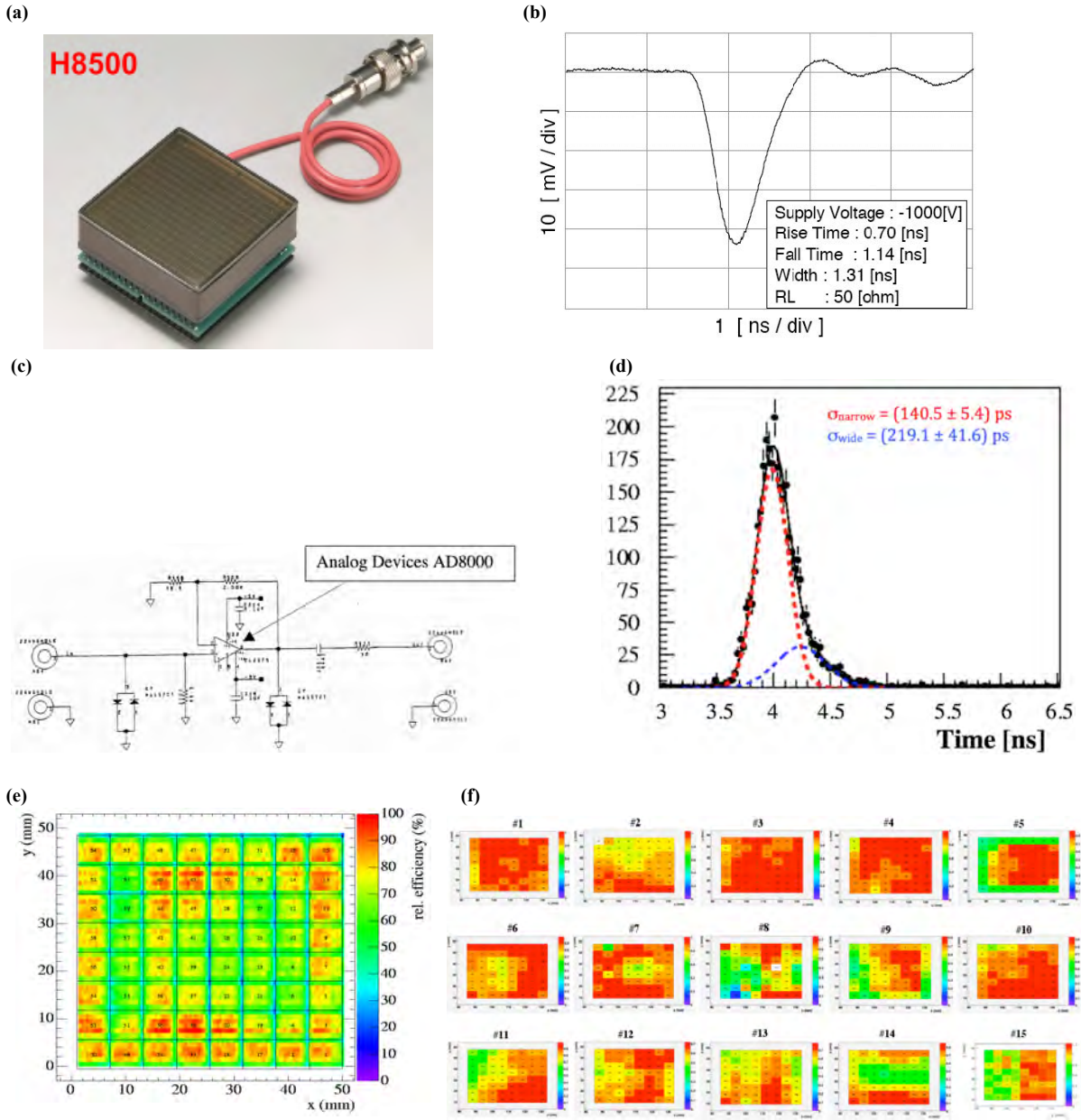

**Fig. 2**: (a) Photograph of an H-8500 MaPMT photodetector; 12 were used in the FDIRC. (b) A typical shape of a signal from an H-8500 without an amplifier. (c) A single-stage SLAC amplifier [6-8]. The early version used the Elantek 2075 chip, the present version is based on the Analog Devices AD8000 chip. (d) An example of the single-photoelectron timing resolution from an H-8500 MaPMT, measured using the SLAC amplifier and the CFD electronics [5]. (e) A detailed single-photoelectron x-y scan across one H-8500 MaPMT at 407 nm, using the SLAC amplifier & CFD electronics. (f) These are single photon relative detection efficiencies of the H-8500 MaPMTs, measured at the center of each pad (there are 64 pads per PMT), and normalized to the detection efficiency of the 2-inch dia. Photonis XP-2262B PMT at a wavelength of 430nm (PiLas laser diode). Electronics in this particular measurement consists of SLAC amplifiers and 32-channel CFDs with 100 mV thresholds followed by Phillips 7188 TDCs. Color coding in this plot: red is 1.0, green is 0.5, yellow is 0.8, where scale 1.0 is the same efficiency as the Photonis tube. This measurement was used to correct for pixel-to-pixel inefficiency in the Monte Carlo (MC) simulation.

## 3. Photon detectors and electronics

The focal plane of the FBLOCK was populated with 12 of the Hamamatsu H-8500 multi-anode photomultiplier tubes (MaPMT) shown in Fig. 2a; this only partially covers the focal plane of the FBLOCK – complete coverage would require 48 such photomultipliers (PMTs). The location of the 12 MaPMTs was chosen to optimize coverage for the expected photon hit locations from "roughly" vertical cosmic ray muons hitting the bars in the SLAC CRT. Each MaPMT provides 64 pixels, each approximately 6×6 mm$^2$. H-8500 PMTs are moderately fast: Fig. 2b shows their rise time of ~0.7 ns, and pulse width of ~1.3 ns. To utilize the speed of the tube, we decided to employ a slower amplifier with a modest gain of 40 [4-6]. Figure 2c shows its circuit. The same amplifier concept was used successfully on the first FDIRC prototype [4-6]. Using this amplifier and constant fraction discriminator (CFD) electronics, we measured a single photo-electron Transit Time Spread ($\sigma_{TTS}$) of ~140 ps [4,5], as shown in Fig. 2d. The H-8500 tube does not have very uniform response; Fig. 2e shows a 2D scan of the single photo-electron response across all 64 pixels using the first FDIRC prototype electronics [4,5]. Figure 2f shows scans for all 12 tubes used in this prototype using the central position of each pixel. These measured parameters are incorporated into our Monte-Carlo simulation and data analysis.

The SLAC amplifier's PC-board had to be modified to accommodate the IRS2 electronics,[2] as shown in Figures 3a&b. Fig. 3c shows the electronics mounted on the FDIRC prototype. A more detailed description of the IRS2 electronics is provided in the next section.

The proposed *SuperB* design, shown in Fig. 3d, utilized CFD-on-a-chip electronics [18]. However, as it was not available before the end of the data taking, the IRS2 was used instead to readout the FDIRC prototype as discussed below.

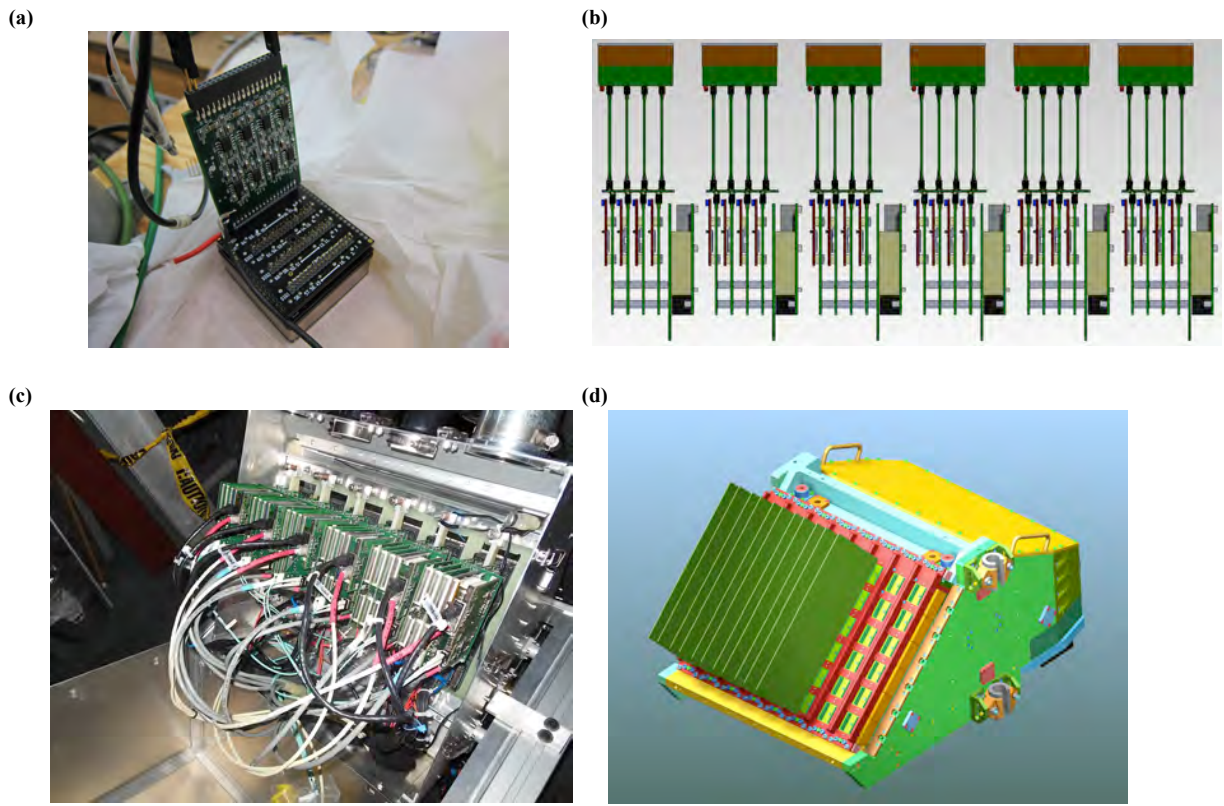

**Fig. 3**: (a) An H-8500 PMT with one amplifier card. (b) A schematic description of detectors, amplifiers and IRS-2 waveform digitizing electronics in this prototype. (c) Final IRS2 and SLAC amplifier electronics mounted on the FDIRC prototype. (d) Planned *SuperB* electronics design to readout all 48 H-8500 MaPMTs of a photon camera coupled to a bar box [1,14].

---

[2] The Ice Radio Sampler (IRS) is an offshoot of the Buffered LABRADOR (BLAB) series of ASICs [19,20], originally designed for performing radio searches for ultra-high energy neutrinos in antennas embedded in Antarctic ice.

# 4. IRS2 electronics

After passing through a SLAC amplifier, the PMT signals are coupled into the IRS2 application-specific integrated circuit (ASIC). The basic architecture of the IRS2 is shown in Fig. 4a.

Each of the 8 input channels couples into a 128-cell switched capacitor array. A pair of externally supplied sampling strobes propagates down a delay line and controls the opening and closing of the switches to the capacitor array. This results in an effective sampling rate 128 times that of the frequency of the sampling strobe. The IRS2 is designed to operate at sampling rates from 1 to 4 gigasamples-per-second (GSa/s). For the FDIRC, the IRS2 is operated at a sampling rate of 2.7 GSa/s. With 128 storage cells, this corresponds to a buffer depth of 47 ns. In order to accommodate higher trigger latencies, the IRS2 contains 32,768 storage capacitors per channel. These storage cells are arranged in 512 blocks of 64-samples each. The sampling array is divided into two 64-sample blocks, each of which connects to 256 of the 512 storage blocks through an array of buffer amplifiers. A 9-bit write address bus and a write strobe are provided externally to choose which storage block to write to and perform the copy operation, respectively. Writing to the storage array is done in "ping-pong" mode: the first 64 stored voltages in the sampling array are copied to a storage block while the second 64 are actively sampling, and the second 64 are copied while the first 64 are actively sampling. As the FDIRC operation requires only a ~2 μs latency, we utilize only the first 64 analog storage blocks. Figure 4b shows a fully assembled module, complete with its amplifier interface card.

In addition to an analog memory capacitor, each storage cell includes a comparator for use in a ramp-compare conversion process. The output of an on-chip analog ramp generator is buffered to each of the storage cells for all channels in parallel. An externally-controlled digitization address bus is used to select a set of 64 comparator outputs from a storage block for connection with an array of 12-bit counters running off of an internal ripple oscillator. Counter values are latched once the ramp exceeds the stored voltages. Typical ramp times are 5-10 μs. Storage blocks are digitized for all channels in parallel. A 9-bit readout address bus (3 bits for channel and 6 bits for sample) is used to provide random access to the digitized voltages.

In order to select a subset of storage windows for digitization, each IRS2 input channel also has a comparator to allow the detection of signals exceeding an input threshold, provided by an external DAC. By monitoring the outputs of these comparators during analog sampling and storage operations, storage blocks can be marked as having a signal of interest to be digitized upon receipt of a system trigger. For operation at the FDIRC, the controlling FPGA reads out 4 storage blocks (256 samples) for each threshold crossing occurring within the event's time window.

Raw IRS2 waveforms require a substantial amount of calibration and processing in order to reduce the full waveform to a measured photon time of arrival. The pulse amplitude is also measured and used primarily to reject noise and waveform artifacts.

Process variations in the sampling and storage cells, (i.e., threshold variations in the ramp-slope comparators) result in an unique ADC code offset to each sampling location. The first step in processing waveforms is to remove these "pedestals". Pedestals are calculated from a special pedestal run taken prior to the beginning of each data taking period. An example of a pedestal-subtracted waveform from PMT signal is shown in Fig. 4c.

The time-of-arrival of each pulse is calculated based on a software CFD method [20]; this method was found competitive with other timing techniques [21]. Linear interpolation is used between points to obtain sub-sample precision on the time of arrival. The amplitude is taken as the maximum point on the waveform.

The layout of the sampling array of the IRS2 causes an asymmetry in the receipt of sampling signals for odd and even cells. This skew causes adjacent odd and even samples to have temporal spacing far from the nominal sampling rate. In some cases a sample further down the array may have an effective sampling time that precedes the sample prior to it. To avoid considerable complications from this effect, timing measurements are performed using only odd samples. This effectively reduces our sampling rate to 1.35 GSa/s.

The sampling strobe of the IRS2 ASIC is provided by an external clock source, which runs asynchronously to the system trigger, and by its nature is stochastic. As such, times calculated from the IRS2 feature extraction must be corrected to account for the phase of the sampling strobe relative to the trigger. This correction is obtained event-by-event by recording the phase of the sampling strobe with a CAMAC TDC. Calculated times are adjusted by this phase measurement before use in Cherenkov analysis.

Although the IRS2 is operated in a continuous sampling mode, regions of interest are only digitized if a channel-level comparator was able to identify a signal over threshold. To set appropriate comparator thresholds, the DAC controlling the threshold input is scanned while the rate of dark counts for each PMT channel is measured. An example scan is shown in Fig. 4d. These hit rates show a very shallow dependence on the trigger threshold all the way into the noise floor of the IRS2, indicating no significant efficiency loss due to the performance of the channel-level comparators of the IRS2.

Timing resolution is evaluated using two datasets: first, test pulses are injected into the signal chain by coupling a signal through the last dynode of the H-8500. The timing of these test pulses is measured using a simple threshold crossing technique on the digitized data. No significant time-walk is expected in this measurement since the pulses have fixed amplitude. This test allows us to measure the timing resolution of the system, excluding most waveform processing limitations. Since these pulses have no intrinsic amplitude variations, we apply a fixed threshold technique to the recorded waveforms to extract their timing. The second dataset used to evaluate timing performance utilizes single photoelectron laser pulses digitized by the PMTs. As the PMTs have significant variation in pulse height, timing of these pulses is calculated based on a software CFD technique. By comparing the timing resolutions obtained in these two samples, we can study the degradations due to pulse height variations and waveform processing limitations.

Resolutions obtained with the fixed-amplitude pulses are in the range of ~100-200 ps. An example of a timing distribution for these signals is shown in Fig. 4e. This includes a detailed calibration for internal ASIC effects and indicates that such resolutions are in principle achievable. However, significant degradation is observed in the laser-induced single photoelectron PMT signals, as shown in Fig. 4f. Typical timing resolutions for these signals are around ~600 ps. The large difference in resolution between these samples suggests that variation in the IRS2 amplitude response is the primary cause of the degradation.

Such amplitude variations arise due to instabilities in operating the IRS2. In particular, there are a large number of control signals required to operate the ASIC, with tight timing margins required between them. Small misalignments of these signals can cause complicated systematic amplitude variations. The IRS2 also has no on-chip method to lock the time-base, (e.g., a DLL or PLL), so these amplitude variations drift with environmental conditions, making it extremely difficult to remove such effects in the full dataset.

A number of revisions have been undertaken to resolve these and other issues with the IRS2 architecture. Notable improvements include the addition of an intermediate storage array that effectively doubles the amount of time available to perform the sensitive analog copying operations, the implementation of an on-ASIC delay locked loop to stabilize the time base, and an internal timing generator that allows very fine internal adjustment of the signals required to perform the analog sampling and copying steps. These improvements have culminated in the IRSX ASIC,[3] which will be used to read out the TOP counter of the Belle-II experiment.

**(a)**

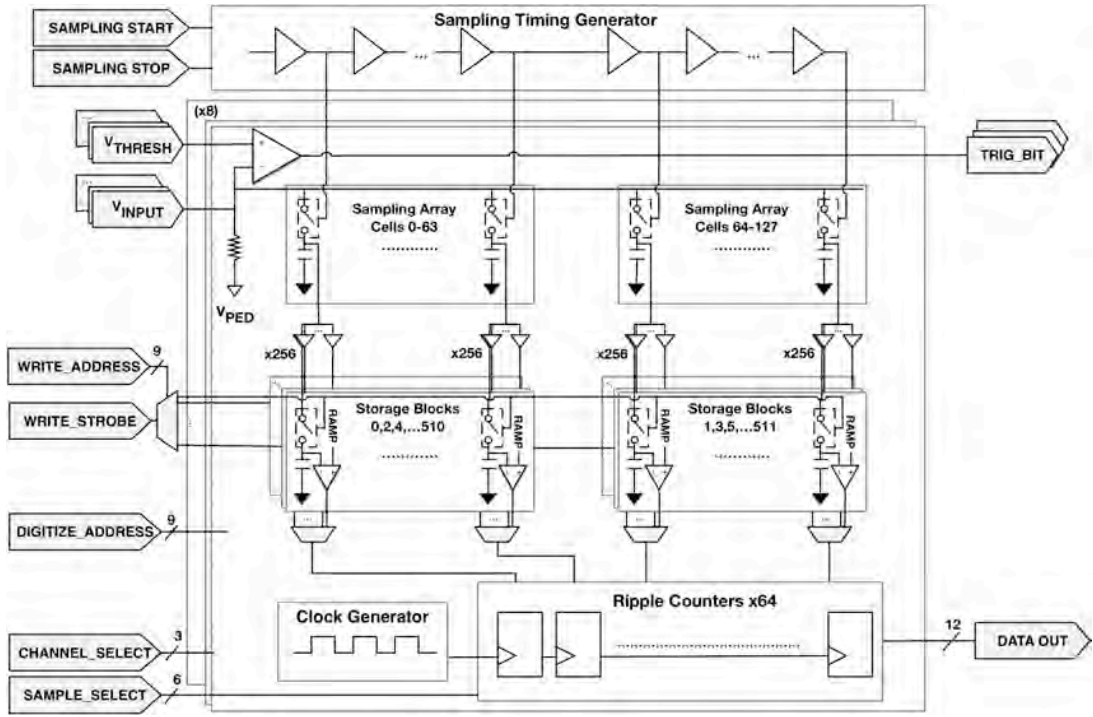

---

[3] Information on the IRSX ASIC is expected to be published in the near future.

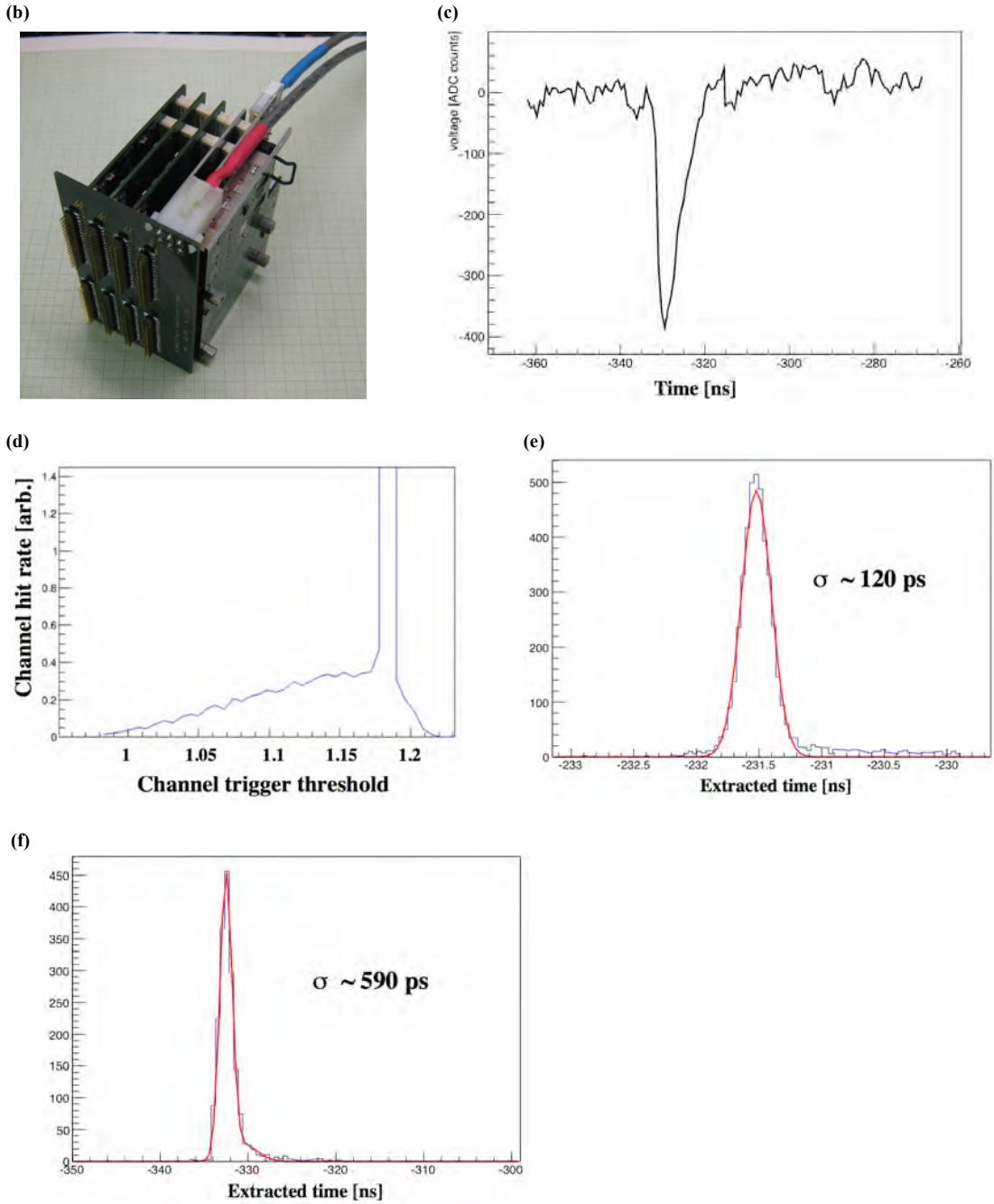

**Fig. 4** IRS2 electronics: (a) Simplified schematic of the IRS2 ASIC. (b) Front view of a fully assembled module, complete with amplifier interface card. (c) An example of a pedestal-subtracted waveform from a PMT signal. (d) Hit rate for an example IRS2 channel as a function of its trigger threshold, as set by a DAC. The nominal voltage offset for each channel is just under 1.2 V. Signals are negative going, so a lower absolute voltage threshold selects a larger pulse. (e) Timing resolution obtained by a pulser signal injected into the H-8500's dynode 12; resolutions obtained with the fixed-amplitude pulses are in the range of ~100-200 ps. (f) An example of a timing resolution with laser-generated single photoelectron signals from an H-8500 MaPMT. Typical timing resolutions for these signals are around ~600 ps.

# 5. Cosmic ray telescope

The FDIRC prototype was tested in the CRT [11]. The telescope features include: two planes of scintillator hodoscopes that provide 3-D tracking of cosmic ray muons with an angular resolution of ~1.5 mrad and track position resolution of 4-5 mm; a fused silica start counter providing start time resolution better than 70 ps; and an instrumented iron range stack used to select muons with energy greater than ~2 GeV. The scintillator hodoscopes had active areas of 51 cm × 107 cm and an angular tracking acceptance of ±17° × ±27° in two respective directions. The bar box was located within the CRT such that the cosmic muons were incident near the midpoint along the length of the fused silica radiators. Figure 5a shows the overall view of the stack of iron and lead, interleaved with large scintillator paddles, tracking hodoscopes, start counter and FDIRC prototype. Figure 5b shows the FDIRC prototype in the CRT, prior to the installation of electronics. Figures 5c and 5d show the start counter consisting of two fused silica bars coupled to a 4-pixel MCP-PMT, as shown in Fig. 5e. Figure 5f shows the 16 × 16 fiber hodoscope with 2 × 2 mm$^2$ scintillating fibers, which was used to measure the tracking resolution.

Figure 6a shows the timing resolution of the start counter obtained in the test beam. However, for the FDIRC data analysis, the start time has to be corrected for the Cherenkov photon TOP within the fused silica bar and for TOF between the start counter and the bar box bar, as shown in Fig. 6b. Figure 6c shows the typical magnitude of this correction, which depends on the track parameters.

The remaining plots of Fig. 6 show the typical ranges of various variables allowed by the CRT within the coordinate system used in the analysis (Fig. 6d). The distribution of the polar and azimuthal track angles are shown in Fig. 6e. Because the start counter was rather large, the resulting footprint of track impact with the bars was not small. Figure 6f shows the x and y track hit footprint in the bar box plane, and in terms of the bars numbers[4]; one can see that roughly 5 out the 12 bars fall within the acceptance of this setup, and contain hits. Figure 6h shows a measurement of the CRT tracking resolution using the fiber hodoscope (detail shown on Fig.6g) as a reference, verifying that we have a 4-5 mm resolution for most tracks. The results to follow are based upon over 600,000 cosmic ray triggers in the CRT.

**(a)**

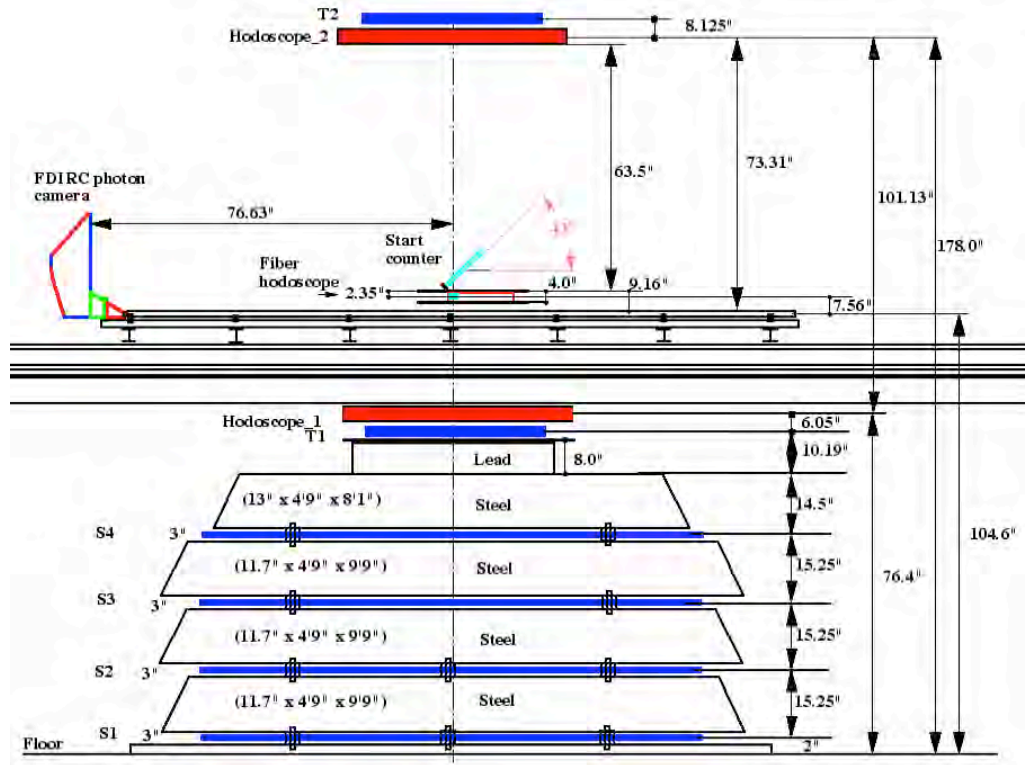

---

[4] In the bar box, the fused silica bars are numbered from 1 to 12 based on their position along the x axis of the analysis coordinate system.

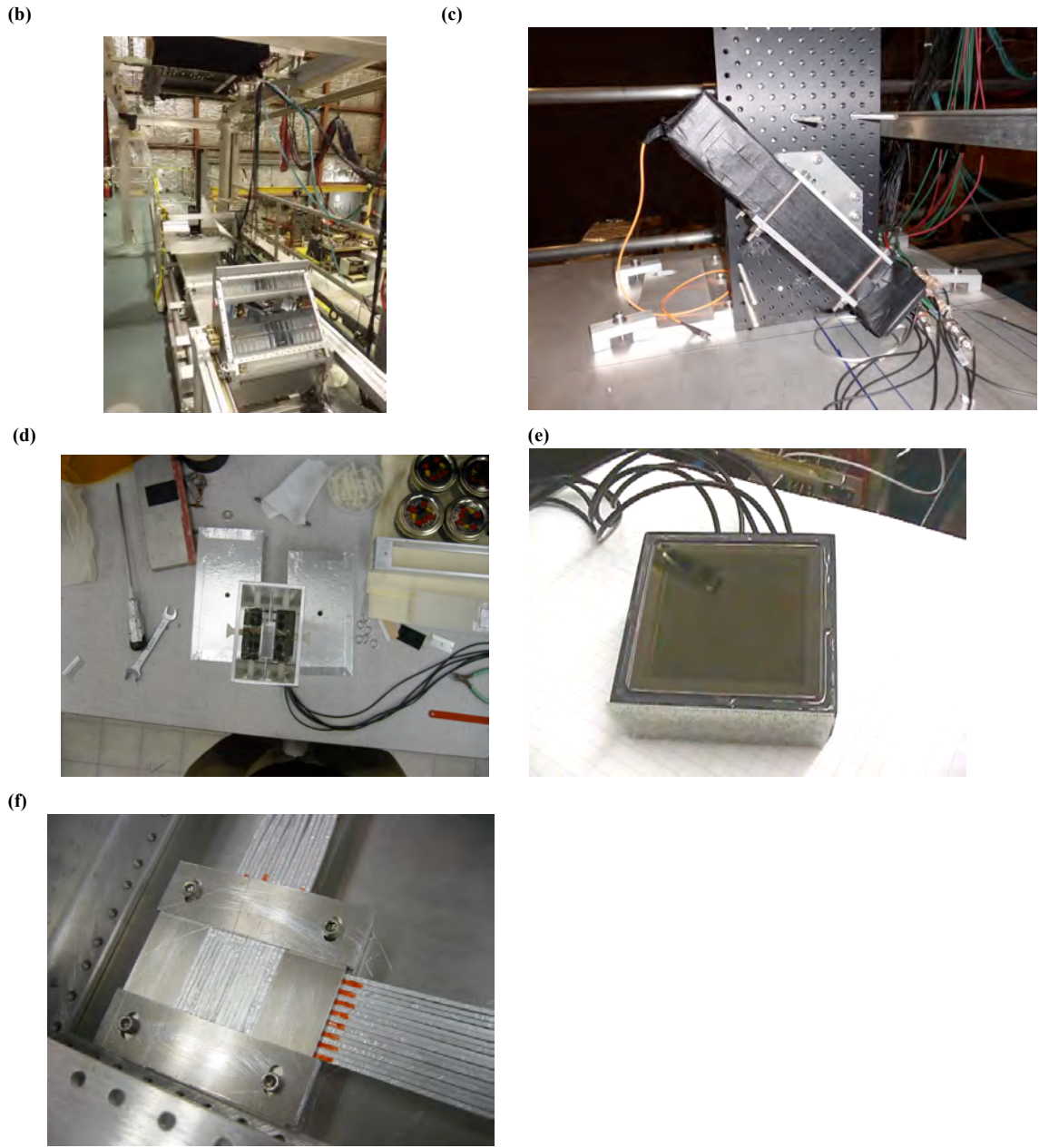

**Fig. 5** Cosmic ray telescope (CRT) [11]: (a) Overall view of stack of iron and lead, interleaved with large scintillator paddles, tracking hodoscopes, start counter and FDIRC prototype. (b) FDIRC prototype in CRT, without electronics at this stage. (c) Start counter consisting of two fused silica bars and MCP-PMT detector. (d) The two start counter fused silica bars viewed from the end. (e) Burle 4-pixel MCP-PMT is an early R&D version, not produced any more. (f) 16x16 fiber hodoscope with $2 \times 2$ mm$^2$ scintillating fibers.

**(a)**

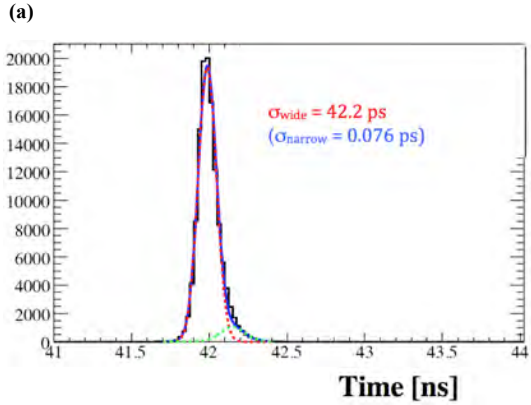

**(b)**

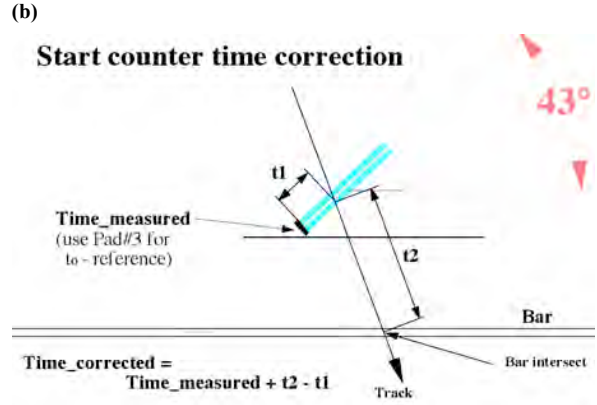

**(c)**

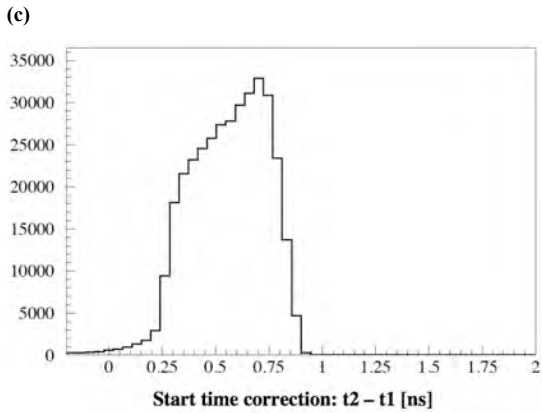

**(d)**

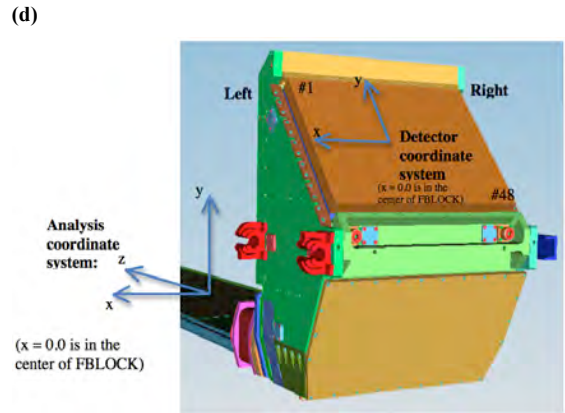

**(e)**

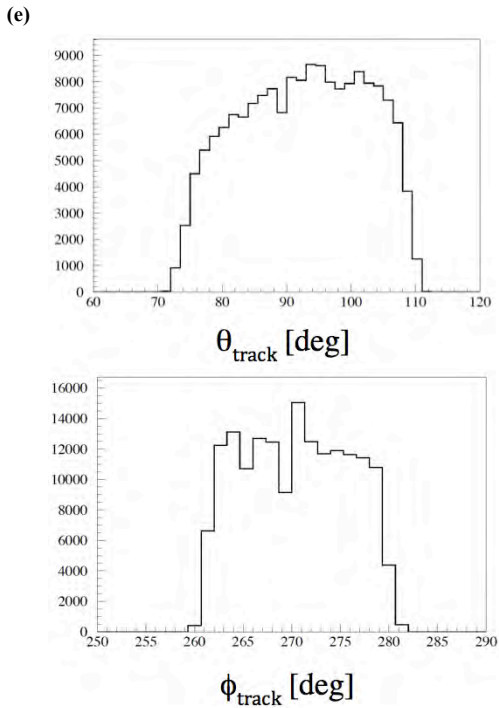

**(f)**

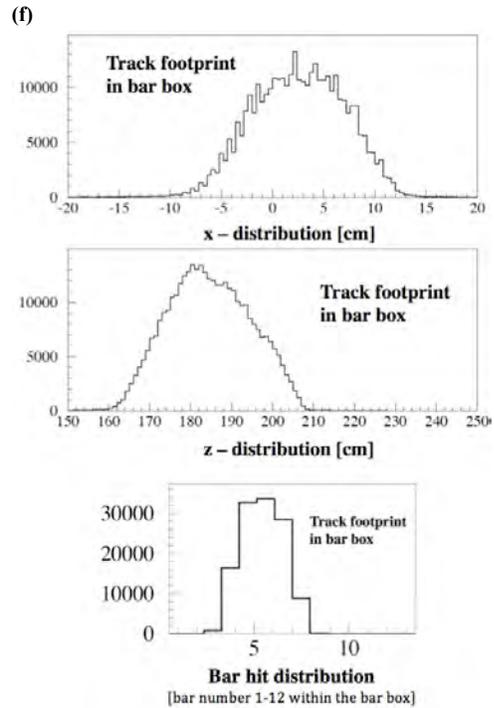

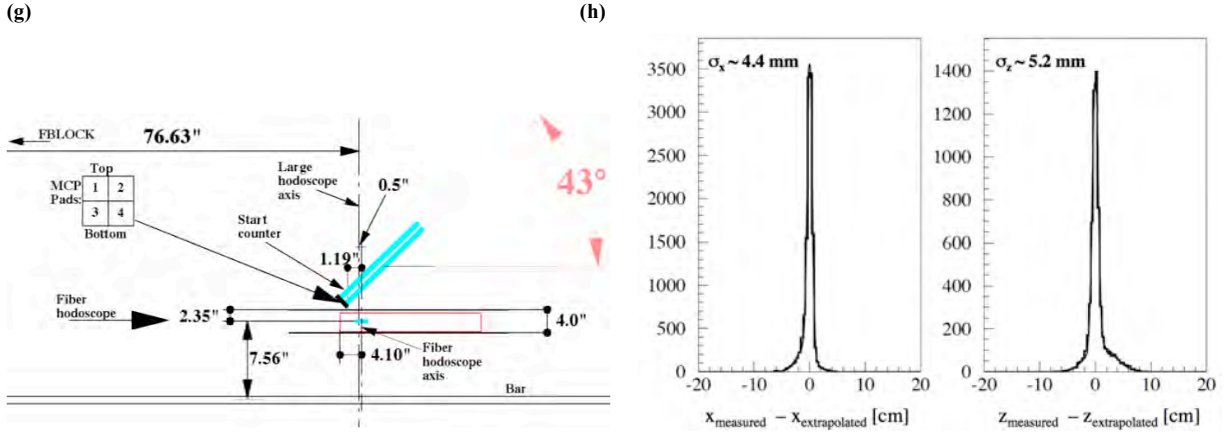

**Fig. 6** Cosmic ray telescope (CRT): (a) Timing resolution of the start counter obtained in the test beam (fitted with two Gaussians). (b) A schematic principle of the start time correction using t2 and t1 times. (c) Typical values of the start time correction. (d) Coordinate system used in the analysis. (e) CRT track distribution in polar and azimuthal angles. (f) x & y track hit footprint in bar box plane, and bar hit footprint: hits from ~5 bars are contributing to this measurement. (g) Geometry of the start counter and the fiber hodoscope within CRT. (h) Measurement of the CRT tracking resolution using the fiber hodoscope as a reference.

## 6. Laser calibration

To calibrate the time offsets for each PMT in the FDIRC, we used laser[5] light striking a diffuser,[6] located at the bottom of the FBLOCK, as shown in Fig. 7a. The diffuser was selected based on uniformity of scattered light, demonstrated visually in Fig. 7b. It provided a relatively uniform spray of photons in the detector plane, although there are some sub-ns time offsets – see Fig. 7c. The intensity was adjusted so that one deals with single photon hits per pixel, and the delay was set so that the laser hits did not overlap with the cosmic ray events, as shown in Fig. 14a in chapter 7.6. In this way we could use the CRT to trigger the laser externally. This enabled continuous real-time calibration, which proved to be very useful to check the stability of the system over many months of running.

Figure 8 shows the pixel timing alignment before and after the laser calibration, when individual pixel offsets were applied. The offset for each pixel was determined using a simple mean of all laser hits within a tight TDC window, and aligning them all to a single value. Fig. 7c shows that in locations where MPTs are located in this prototype, there is a small spread of relative time offsets of less than 200 ps, justifying the alignment procedure of Fig. 8, as long as the electronics resolution is ~600 ps.

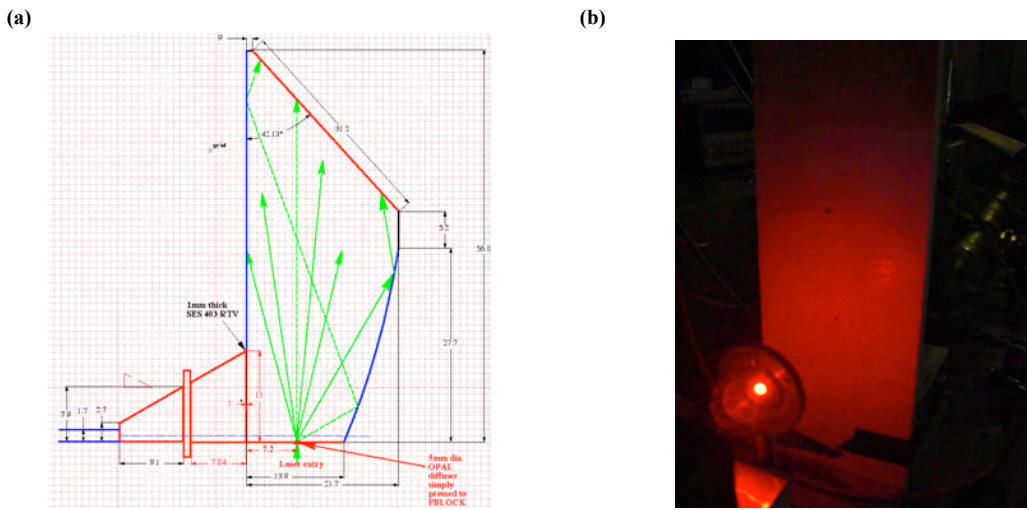

(a)  (b)

---

[5] PiLas laser, 407 nm, made by Advanced Laser Diode Systems, D-12489 Berlin, Germany
[6] Opal diffuser, 5 mm dia., P/N K46-162, Edmund Scientific

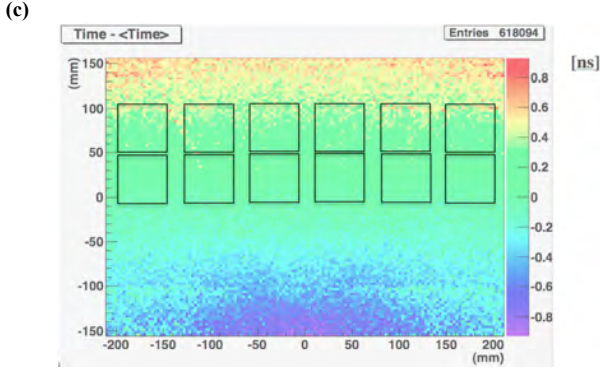

**Fig. 7** Laser calibration: (a) Laser entry point in FBLOCK. (b) Laser light scatters randomly from the diffuser. (c) MC simulation of relative time offsets of the laser light on the detector focal plane. The picture shows the PMT locations.

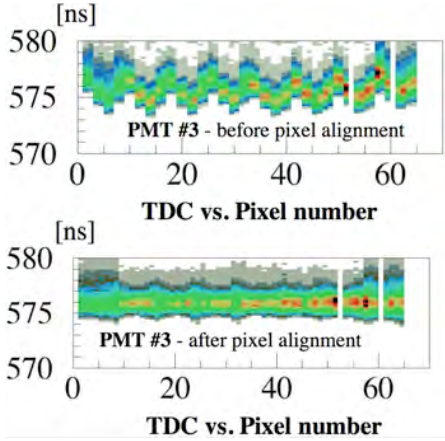

**Fig. 8** Pixel alignment before and after laser calibration.

## 7. Experimental results

To reconstruct the Cherenkov angle in a given event, we used a MC-generated dictionary that maps each MaPMT pixel to a photon direction inside the fused silica bar, $\mathbf{k}_{pixel}$, as well as the photon TOP inside the FBLOCK [22]. The dictionary was generated by simulating propagation of single photons in the fused silica bars, drawn from an isotropic initial angular distribution inside the bar. There may be more than one photon path that will lead from the fused silica bar to a given pixel, giving rise to multiple solutions of $\mathbf{k}^{pixel}$ for a given MaPMT hit. The dictionary is therefore multi-valued [23]. Many of these ambiguous solutions can be eliminated by a cut on dTOP = (TOP$_{measured}$ - TOP$_{expected}$). In addition, hits had to satisfy the internal reflection law on the FDIRC optical surfaces, have real physical values, i.e., a reasonable value of ADC, TDC, not too many hits per single PMT (corresponding to possible cross-talk), and pulses produced by waveform digitizing electronics had to be above a given threshold. Figure 9a shows an example of such a correlation between dTOP and $\theta_c$, showing many well-separated unphysical ambiguity solutions for backward[7]-going photons (the correct solutions are near dTOP ~0 ns and $\theta_c$ ~825 mrad).

The cosine of the Cherenkov angle was then calculated as the dot product of the track direction (measured by the CRT) and the photon direction:

$$\cos \theta_c = \mathbf{k}_{track} \cdot \mathbf{k}^{pixel}.$$

---

[7] Backward photon: $k_z^{pixel} > 0$, forward photons: $k_z^{pixel} < 0$.

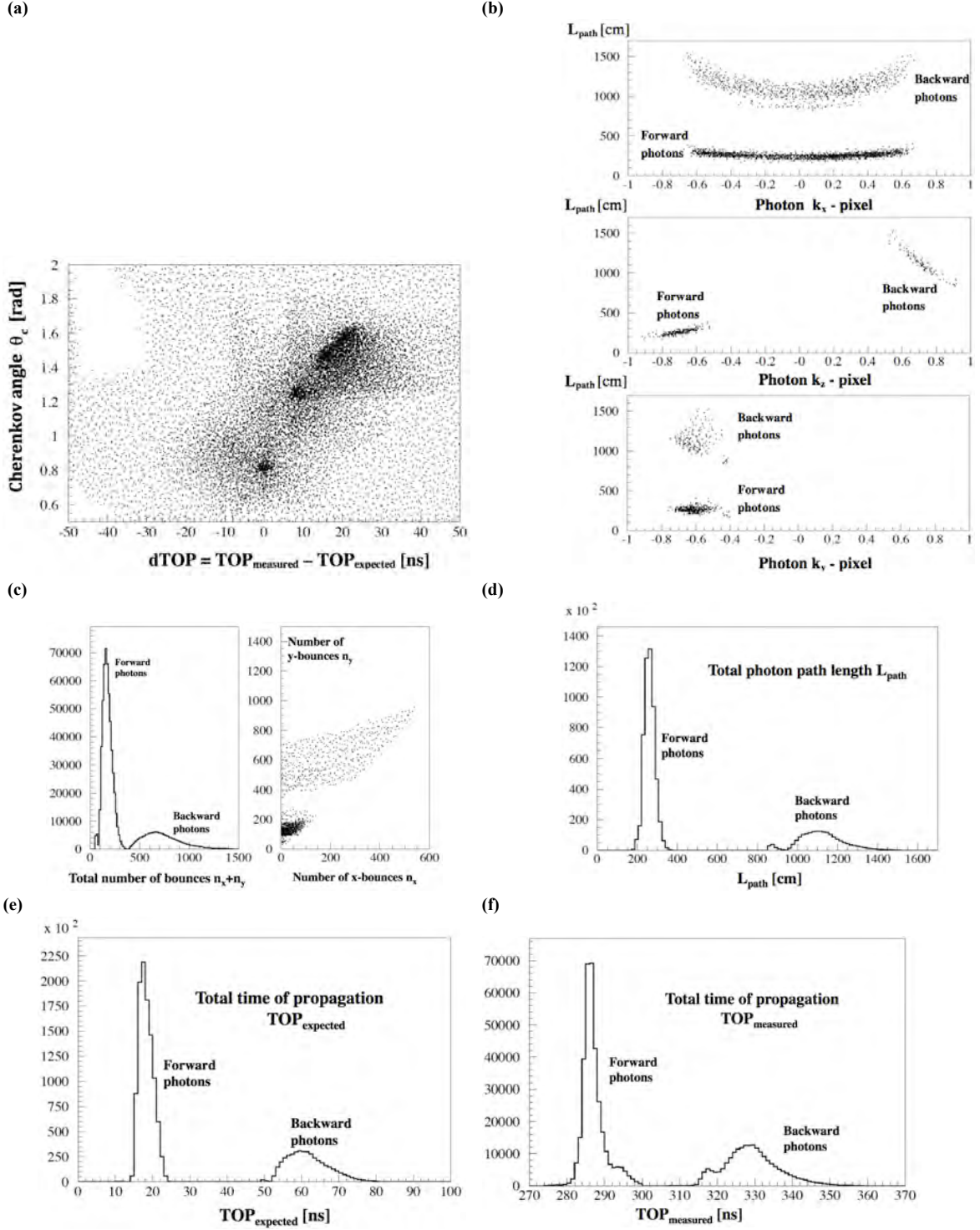

**Fig. 9** (a) Correlation between dTOP and $\theta_c$, showing many unphysical ambiguity solutions for the backward going photons. (b) Photon path length as a function of $\mathbf{k}^{pixel}$ vector components. (c) Total number of bounces in the bar and correlation between x & y bounces. (d) Total photon path length. (e) Calculated total photon time-of-propagation $TOP_{expected}$, which includes time spent in the bar and in the FBLOCK. (f) Measured total photon time-of-propagation $TOP_{measured}$ determined by IRS2 electronics; one has to subtract an arbitrary time offset to match the $TOP_{expected}$ scale.

## 7.1. Range of various variables

The photon propagates in the fused silica with the velocity:

$$v_{group} = c/n_{group} = c/[n_{phase} - \lambda \, dn_{phase}/d\lambda],$$

where $n_{group}$ is the group refraction index, which differs from the phase refraction index, $n_{phase}$, by a few percent. The TOP of a Cherenkov photon in the fused silica bar is then:

$$TOP = L_{path}/v_{group} = L_{path} \, [n_{phase} - \lambda \, dn_{phase}/d\lambda]/c.$$

Figure 9 shows typical ranges of some critical variables for the FDIRC data analysis, such as $\mathbf{k}^{pixel} = (k_x^{pixel}, k_y^{pixel}, k_z^{pixel})$ vector; numbers of photon bounces in the bars, $n_x$ and $n_y$; typical photon path length, $L_{path}$; and measured and expected total photon TOP. The reach of these variables is different than in the case of the first FDIRC prototype [12]. The present FDIRC has a full size bar box, as shown in Fig. 1g, which is longer than the first prototype and has 12 bars compared to the first prototype's single bar. $TOP_{expected}$ is calculated for each PMT pixel as follows:

$$TOP_{forward} = TOP_{FBLOCK} + L_{path-forward} (n_{group}/c)$$
$$TOP_{backward} = TOP_{FBLOCK} + L_{path-backward} (n_{group}/c)$$
$$L_{path-forward} = z_{bar\_entry} /|k_z^{pixel}|$$
$$L_{path-backward} = (2 L_{bar} - z_{bar\_entry})/|k_z^{pixel}|$$

where $L_{path}$ is the photon path length for forward and backward photons, $L_{bar} = 4900$ mm is the total bar length, $z_{bar\_entry}$ is obtained from the CRT tracking, and $TOP_{FBLOCK}$ and $\mathbf{k}^{pixel}$ constants are generated by the MC simulation. Figure 9e shows that it takes typically ~20 ns for forward photons, and ~60 ns for backward photons. For comparison, Fig. 9f shows $TOP_{measured}$ using FDIRC's time measurement. Figure 9d shows the calculated photon path length. Forward photons travel typically 2.5 meters, while backward photons can travel up to ~15 meters inside the bar. The number of photon internal reflection bounces inside the bar is calculated as follows:

$$n_x = L_{path}/(bar_{width} \, |k_z^{pixel}/k_x^{pixel}|)$$
$$n_y = L_{path}/(bar_{thickness} \, |k_z^{pixel}/k_y^{pixel}|)$$

## 7.2. dTOP resolution

The expected $TOP_{expected}$ of a photon can be calculated from the track entry location in the fused silica bar as measured by the CRT, the photon direction $\mathbf{k}^{pixel}$, and the TOP inside the FBLOCK taken from the MC dictionary. This quantity was then compared to the measured TOP to calculate dTOP. This variable provides the primary background rejection and ambiguity resolving power. Figures 10a-d show the measured and simulated dTOP for forward traveling (direct) and backward traveling (indirect) photons. All plots on Fig.10 do have a tight cut on the Cherenkov angle of $|d\theta_c| < 30$ mrad ($d\theta_c = \theta_c - \theta_0$, where $\theta_0 = 825$ mrad is the mean Cherenkov angle in fused silica); this is in addition to all other cuts mentioned earlier in the chapter. The dTOP resolution consists of the following contributions:

$$\sigma_{dTOP} \sim \sqrt{(\sigma_{TTS}^2 + \sigma_{Chromatic}^2 + \sigma_{Electronics}^2 + \sigma_{Pixel\_contribution}^2 + \sigma_{Tracking\_contribution}^2)},$$

where $\sigma_{TTS}$ ~140 ps is the intrinsic H-8500 MaPMT transit time spread (see Fig. 2e), $\sigma_{Chromatic}$ ~ 60-70 ps/m is the photon chromatic broadening per meter of photon path length, $\sigma_{Electronics}$ ~600 ps dominates the dTOP resolution, $\sigma_{Pixel\_contribution}$ and $\sigma_{Tracking\_contribution}$ are negligible compared to the electronics contribution. Resolutions shown on Figs. 10c&d would change to 0.325 ns and 1.03 ns if one switches the electronics resolution contribution off in the MC simulation. The difference in dTOP resolution between forward and backward photons is explained by the chromatic dispersion. Figures 10b and 10d shows that data and MC simulation agree as far as the degradation of the dTOP resolution for backward photons. Figure 10e&f shows the measured dTOP/$L_{path}$ distribution for forward and backward photons.

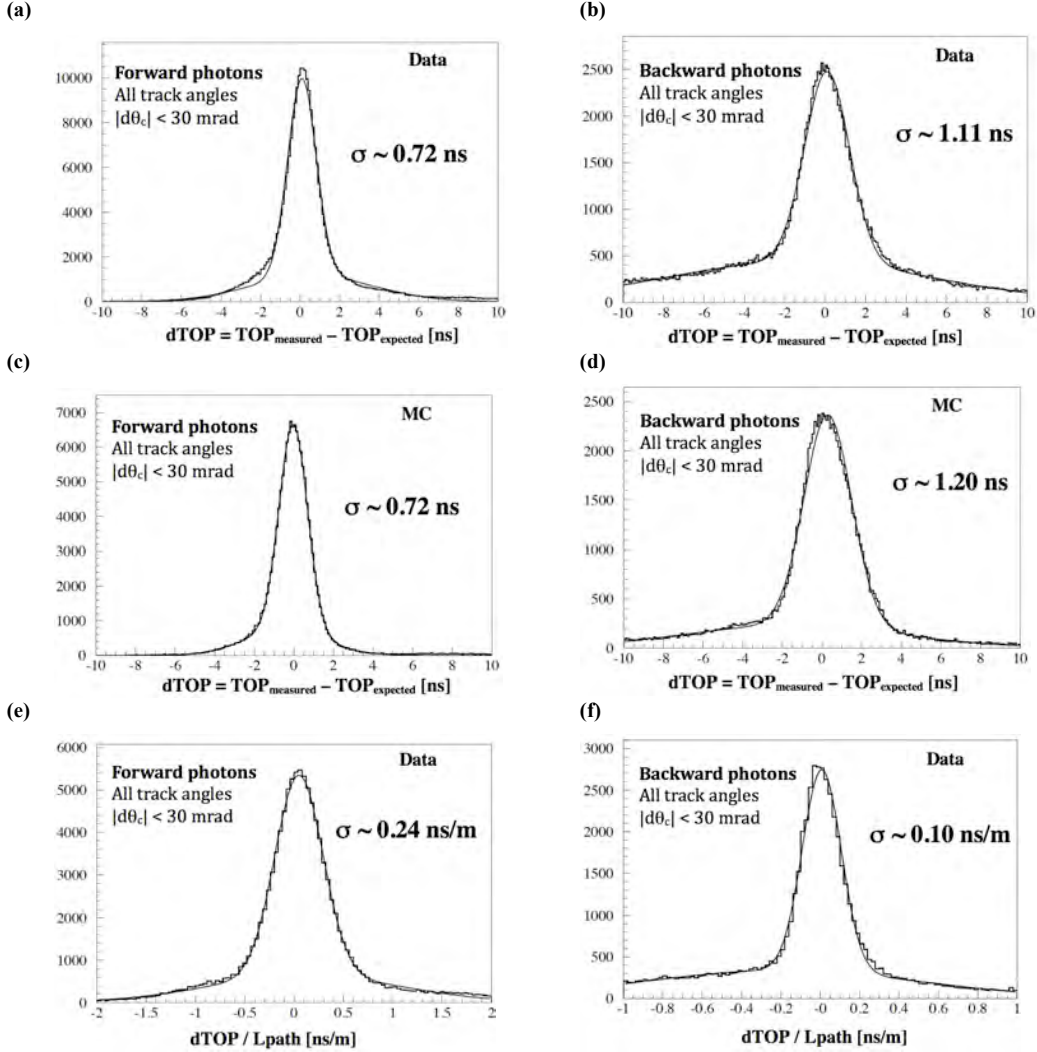

**Fig. 10** Measured dTOP distribution for (a) forward photons and (b) backward photons. Each fit uses a double-Gaussian function, and we quote the narrow one. dTOP MC simulation for (c) forward photons and (d) backward photons. (e&f) Measured dTOP/$L_{path}$ distribution for forward and backward photons.

### 7.3. Cherenkov angle resolution

Figure 11 shows the Cherenkov angle distribution for all photons after the cut |dTOP| < 2 ns for forward photons and |dTOP| < 2.5 ns for backward photons. The measured resolution of 10.4 mrad in Fig. 11a, which includes all track angles and all photons, is somewhat worse than our present MC simulation's prediction of 9.1 mrad. The background under the peak comes primarily from ambiguous solutions to the photon direction, which results from multiple possible photon paths to a given pixel due to reflections on the wedge and FBLOCK surfaces.

The best measured Cherenkov angle corresponds to the very central region of the Cherenkov ring. This is due to the so called kaleidoscopic effect discovered during the first FDIRC prototype R&D effort [6], and later further studied in more detail [24,25]. This optical aberration ranges from 0 mrad (at ring center) to ~9 mrad (in the outer wings of Cherenkov ring), i.e., it is a significant effect, comparable to the chromatic error in magnitude. Selecting backward photons with a small number of bounces ($n_x$ < 10) gives access to the best measured central part of the ring. Figure 11e shows the corresponding Cherenkov angle resolution.

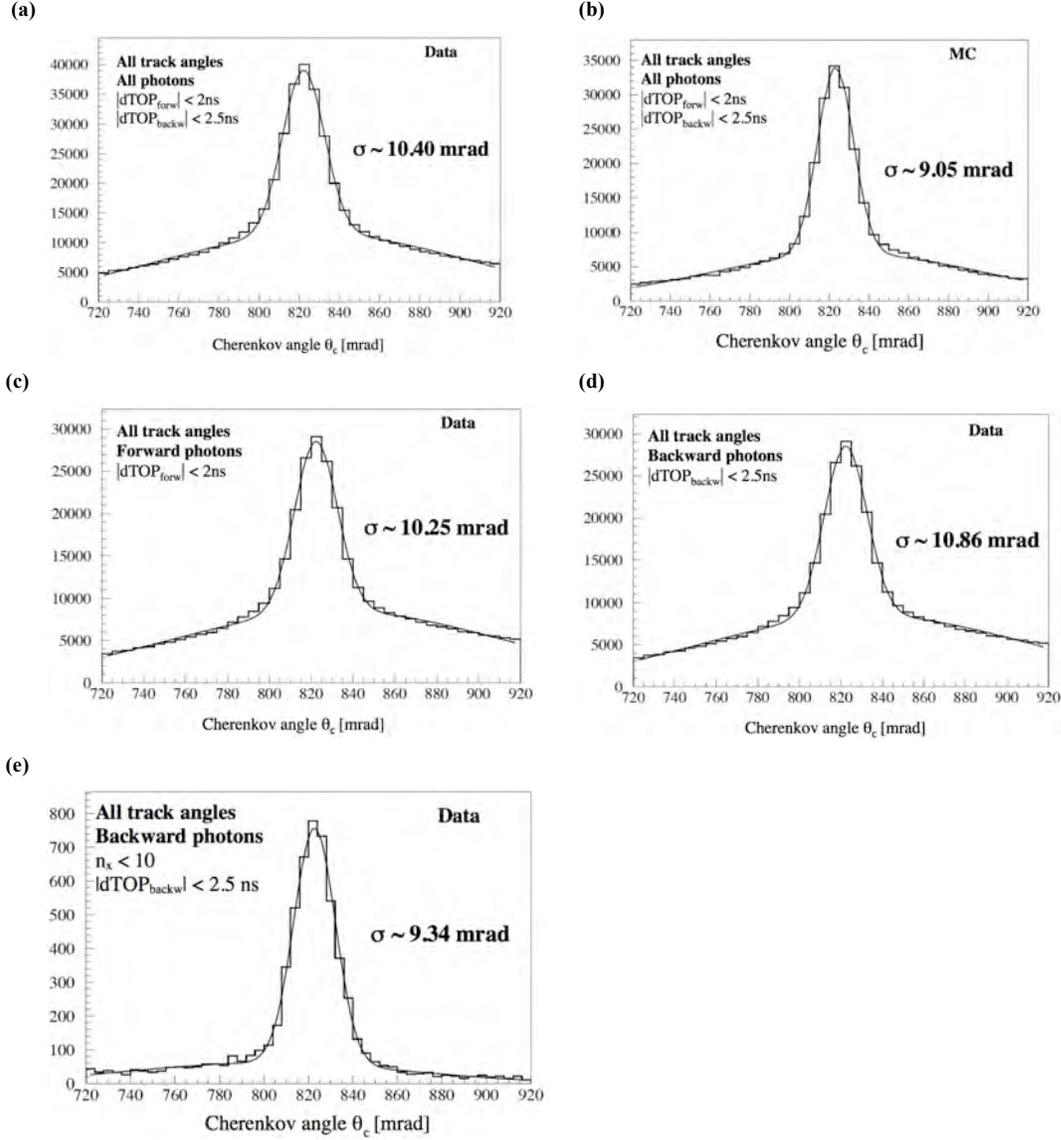

**Fig. 11** Cherenkov angles with dTOP cuts. (a) Data for all tracks and photons; (b) MC simulation of condition (a); (c) All track angles with forward photons; (d) All track angles with backward photons; (e) Photons restricted so that number of bounces is small ($n_x < 10$) and using backward photons only, which restricts the photon to the very central region of the Cherenkov ring. This region provides the best measured Cherenkov angle.

## 7.4. Chromatic correction

The refractive index of the radiator is a function of wavelength. This produces dispersion in the Cherenkov angle since red photons correspond to smaller angles than blue photons. However, the red photons have a larger group velocity and therefore arrive at the detector before the blue photons, according to equation:

$$dt = L_{path}\, \lambda\, d\lambda/c \times (- d^2 n_{phase}/d\lambda^2),$$

where $dt$ is time dispersion, $L_{path}$ is bar length, $n_{phase}$ is a phase refraction index, $c$ is velocity of light, and $d\lambda$ is a wavelength bandwidth. One should point out that the most powerful method to utilize all this information is to carry out a full 3D likelihood analysis, but in order to demonstrate the benefits of the chromatic correction, we will use a different, more direct analysis with these data. Since the $\theta_c$ measured for each photon depends on the wavelength of that particular photon through the Cherenkov equation, by measuring the correlation between $d\theta_c$, which is the difference between measured Cherenkov angle and its nominal value for 410 nm photon wavelength, and $dTOP/L_{path}$, where $L_{path}$ is the total path length of the photon in the detector, one can infer the color of the photon and apply a correction to the measured Cherenkov angle. This is shown schematically on Fig. 12a. Figure 12b shows the actual measurement of this correlation in the FDIRC prototype using 3D tracks, and Fig. 12c shows the corresponding MC simulation. Entries into Fig. 12b are single photons, which had to satisfy the ambiguity treatment mentioned earlier. Using these 2D-maps to measure linear correlations between $d\theta_c$ and $dTOP/L_{path}$, one can then apply a correction to the measured Cherenkov angle as follows:

$$\theta_c^{corr} = \theta_c - \text{Slope} \times (dTOP/L_{path})$$

As the forward photons travel a shorter distance, given our timing resolution, their chromatic correction is less effective. After applying the correction for the backward going photons only, we see an improvement in the Cherenkov angle resolution of approximately ~0.7 mrad, compared to the MC expected improvement of ~1 mrad – see Figs. 12d&e. Figures 12e&f show the same plots for data and MC respectively, but for the best-measured part of the Cherenkov ring, i.e., the central portion of the ring, a condition obtained in this case for backward photons with a number of photon bounces limited to $n_x < 10$.

Figure 12j shows an attempt to correct dTOP distribution for the chromatic broadening using the Cherenkov angle measurement. Comparing this result with Fig.10b, one can see a modest ~ 5.4% improvement (this is to be compared to ~ 6.4% improvement in the Cherenkov angle resolution when doing the chromatic correction). Figure 12k shows the dTOP distribution for backward photons with small number of bounces in x-direction ($n_x < 10$), and chromatically corrected with the Cherenkov angle. This distribution is 5.6% narrower than uncorrected distribution.

**(a)**

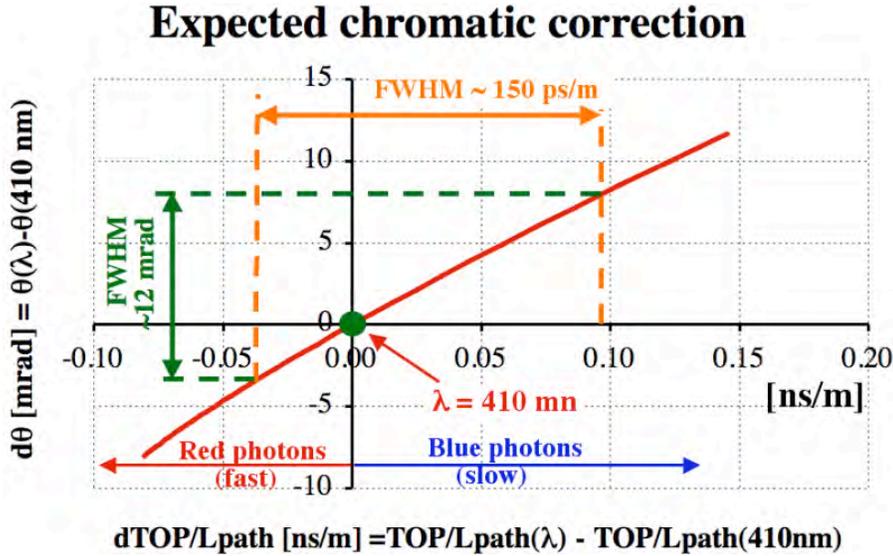

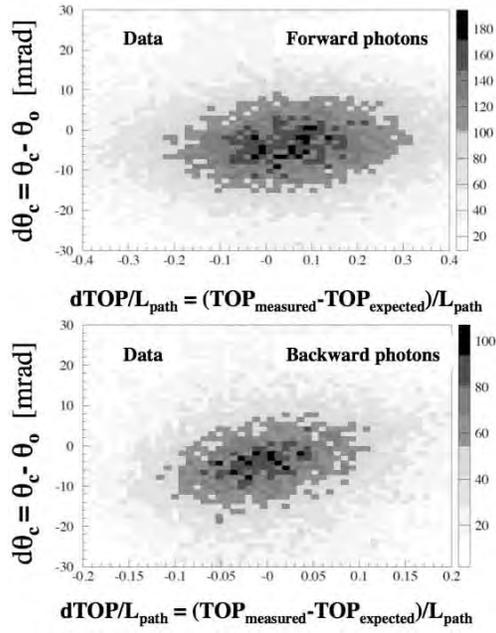
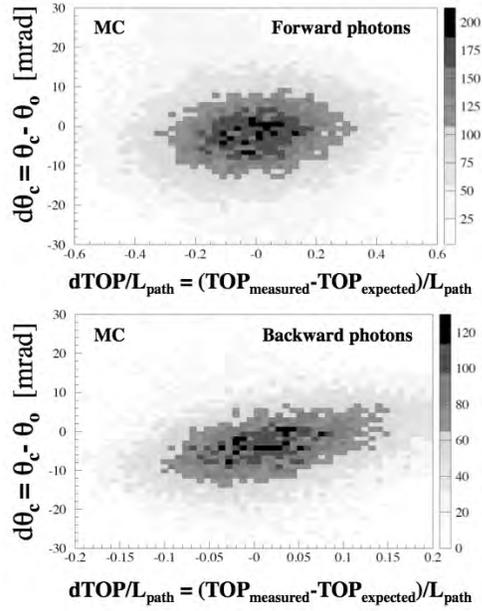
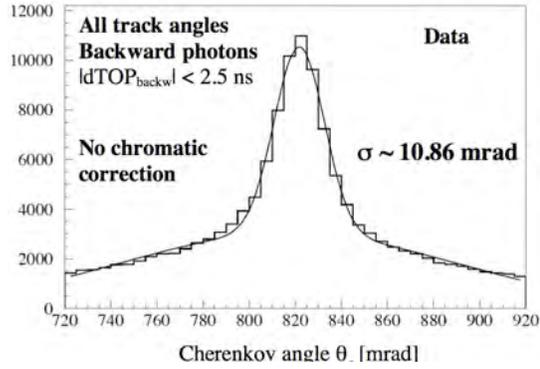
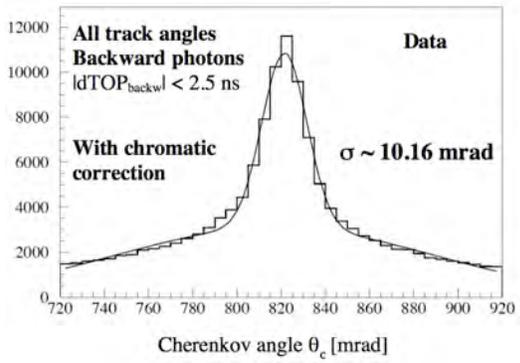
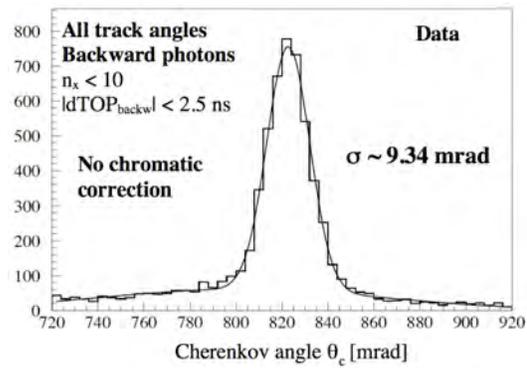
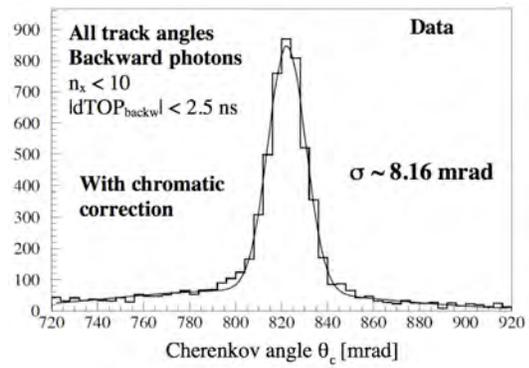

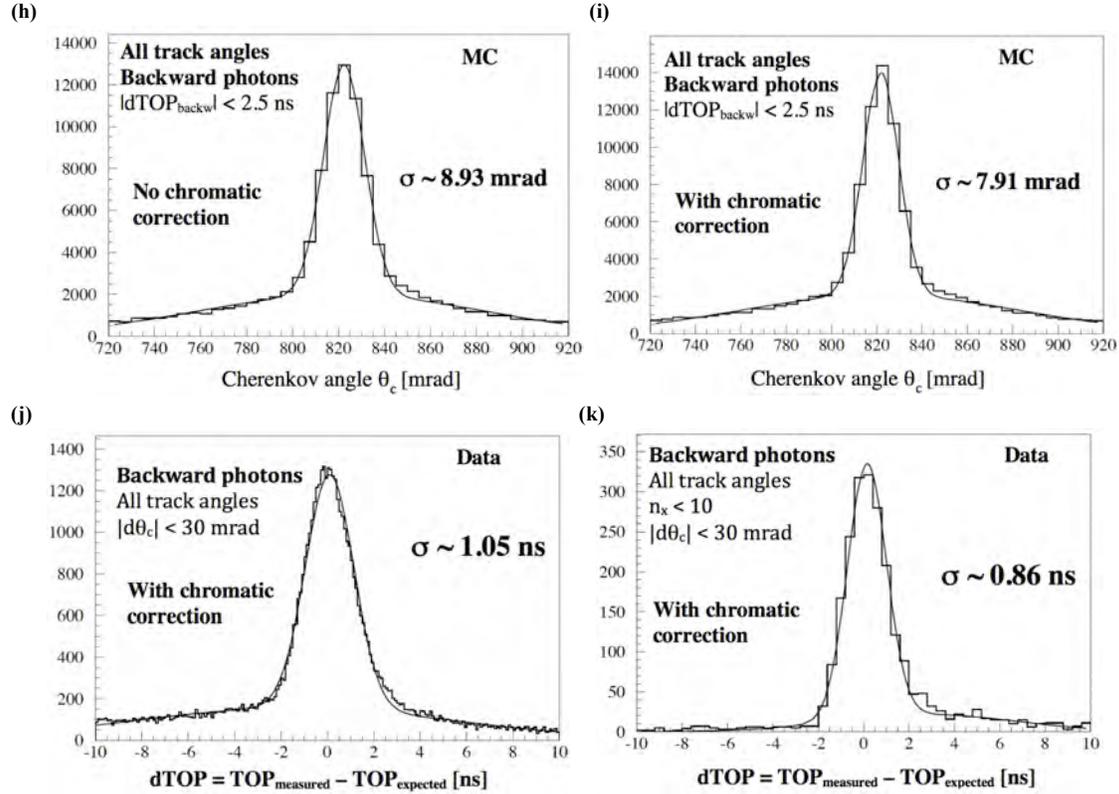

**Fig. 12** (a) A schematic plot of the correlation between $d\theta_C$ and $dTOP/L_{path}$ used to correct the chromatic dispersion of the Cherenkov angle. (b) As shown in (a), for data. (c) As shown in (a) and (b), but for the MC simulation. (d)&(e) Measured effect of the chromatic dispersion correction on the Cherenkov angle for all tracks, for forward and backward photons. (f)&(g) As for (d)&(e), but using only photons in the best measured part of the Cherenkov ring (backward photons with a small number of x bounces, nx <10). (h)&(i) As (f)&(g), but for the MC simulation. (j) dTOP distribution, chromatically corrected with the Cherenkov angle (to be compared with uncorrected distribution of Fig.10b). (k) dTOP distribution for backward photons only and small number of bounces in x-direction ($n_x < 10$), chromatically corrected with the Cherenkov angle. This distribution is 5.6% narrower than uncorrected distribution.

### 7.5. Number of photoelectrons per ring

The number of photoelectrons per track was studied by choosing tracks perpendicular to bars within ± 3°, matching the best CRT coverage. Due to the limited MaPMT coverage of the focal plane, we did not expect a large number of hits per ring. Figures 13a&b show the measurement and the corresponding MC simulation (for the present partially instrumented focal plane). A slight discrepancy can be explained by the following effects : (a) the MC simulation does not describe correctly all detector defects, (b) the detector pixel detection efficiency was measured only at the pixel's center and using the old SLAC electronics, as used in Ref. [4,5], (c) the very first DIRC bar box, which we are using in this prototype, was built using somewhat worse bars, and (d) that bar box was not protected by a flow of boil-off $N_2$ gas while sitting in the clean room for ~15 years. Therefore we consider the agreement as reasonable.

Figure 13c shows the MC simulation of the number of photons per ring for perpendicular tracks, assuming a fully instrumented FDIRC prototype with all 48 H-8500 MaPMTs, each with 90% pad relative detection efficiency, and using a comparable dTOP cut. The relative pad efficiency is normalized to a nominal bialkali QE shape.

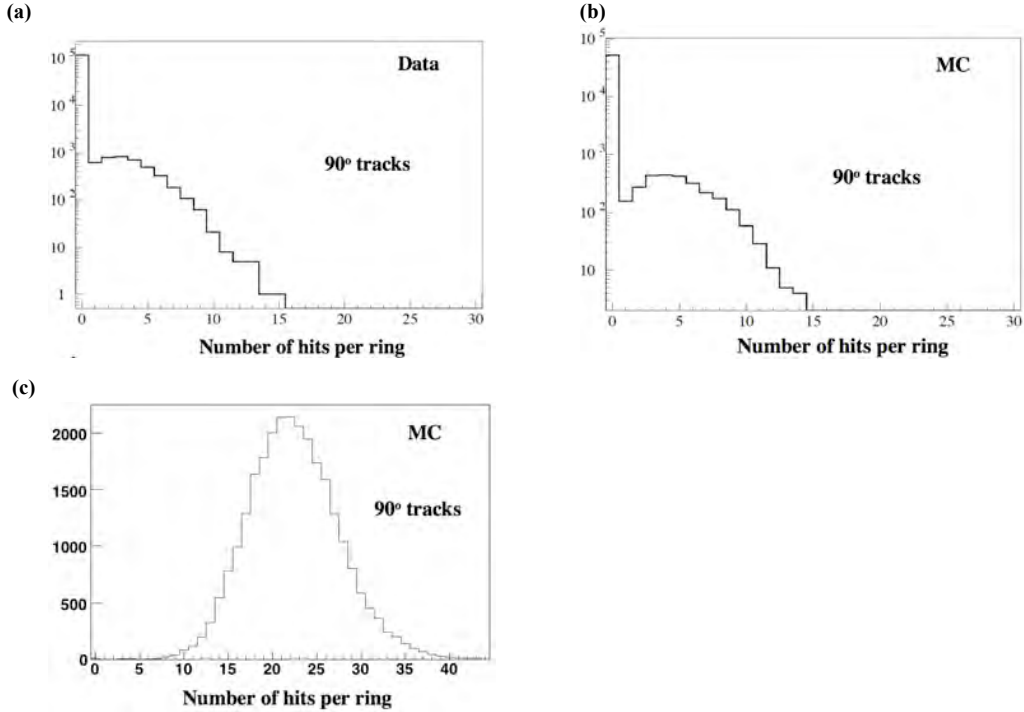

**Fig. 13** (a) Measured number of hits per ring for 90 ± 3° tracks from data. (b) As in (a), but for MC simulated events. (c) MC simulation of number of photoelectrons per ring for 90 ± 3° tracks, assuming a fully instrumented FDIRC with all 48 H-8500 MaPMTs with 90% pad relative detection efficiency, and using the present dTOP cut. The relative pad efficiency is normalized to a nominal bialkali QE shape.

### 7.6. Performance at high rates

To study the performance of the FDIRC detector in the presence of large background, we added a random light source to the usual cosmic ray data stream. The background was generated as follows: we used a fiber splitter[8], where one branch was connected to the PiLas 407 nm laser used for timing calibration, and the other branch to a random light source,[9] coupled to a 400 μm diameter fiber through a lens. The intensity was simply adjusted by Mylar attenuators located between the halogen lamp and the fiber. In the so-called high rate run discussed below (Figs.14 and 15), the random rate was set to ~3 MHz/PMT, or an equivalent background photon rate of ~36 MHz/12 PMTs; this corresponds to 144 MHz/bar box with a fully instrumented FBLOCK with 48 PMTs.

Figures 14a&b show a storage scope display of one H-8500 pixel, showing an accumulated time distribution of the Cherenkov peak (on the left), the laser calibration peak (on the right), together with the trigger on the bottom trace, without and with the random background added.

Figures 15a&b show corresponding TDC distributions with forward and backward Cherenkov photons, and Figs. 15c&d show dTOP distributions; the Signal/Noise (S/N) ratio is significantly worse in the background run in both cases.

Figures 15e&f show that we still see a clear Cherenkov photon peak at the highest background rate. The Cherenkov peak resolution remains virtually unchanged but does sit on a larger background; the S/N ratio goes from ~2.4 at nominal "clean" running conditions to ~1.4 at high background photon rates. In both cases we applied dTOP cuts to reject the background: $|dTOP_{forward}| < 2.0$ ns and $|dTOP_{backward}| < 2.5$ ns.[10] Figures 15g&h show the Cherenkov angle resolution for the best measured region of the Cherenkov ring where the angle resolution and the S/N ratio are the best.

---

[8] Fiber splitter WD202A2-FC made by Thorlabs Co.
[9] SL1-Filter Tungsten-Halogen random light source, 5W, 300-1700 nm, made by Stellar Net Inc.
[10] With better electronics, providing ~200 ps timing resolution per photon, a tighter dTOP cut could be made, providing even better S/N ratio performance.

Our artificial maximum background rate of ~3 MHz/PMT, or ~144 MHz/bar box, corresponds to roughly one and a half times what the Belle-II TOP counter is predicting as their maximum random rate at Super-KEKB,[11] which is ~3 MHz/PMT, or ~96 MHz/bar box. For event reconstruction, it is important to consider the total number of pixels available, both in space and time domains. Assuming that the "time domain pixel" is equal to the "time resolution" for simplicity, and the total extent of one event equal to ~70 ns, the FDIRC has 48 × 64 × 350 ~1.1 million pixels (350 = 70 ns/200 ps).

For PMT aging, the important quantity is the total accumulated charge in Coulombs/cm$^2$, which turns out to be 3-5 C/cm$^2$ for the FDIRC, assuming a total PMT gain of $10^6$ and a 10 year-long exposure; this would result in a negligible loss of gain, according to Hamamatsu data.

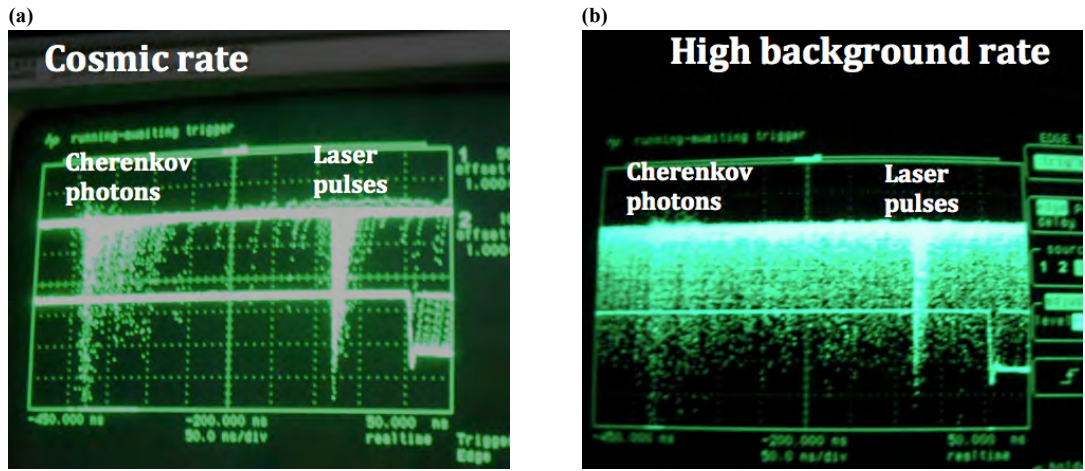

**Fig. 14** Scope display of one pixel for (a) Cherenkov and calibration laser photons during a normal CRT run. Bottom trace shows CRT trigger. (b) As (a) but with an overlay of random background from the lamp.

---

[11] Private communication with Prof. Kenji Inami, Nagoya Univ., Japan.

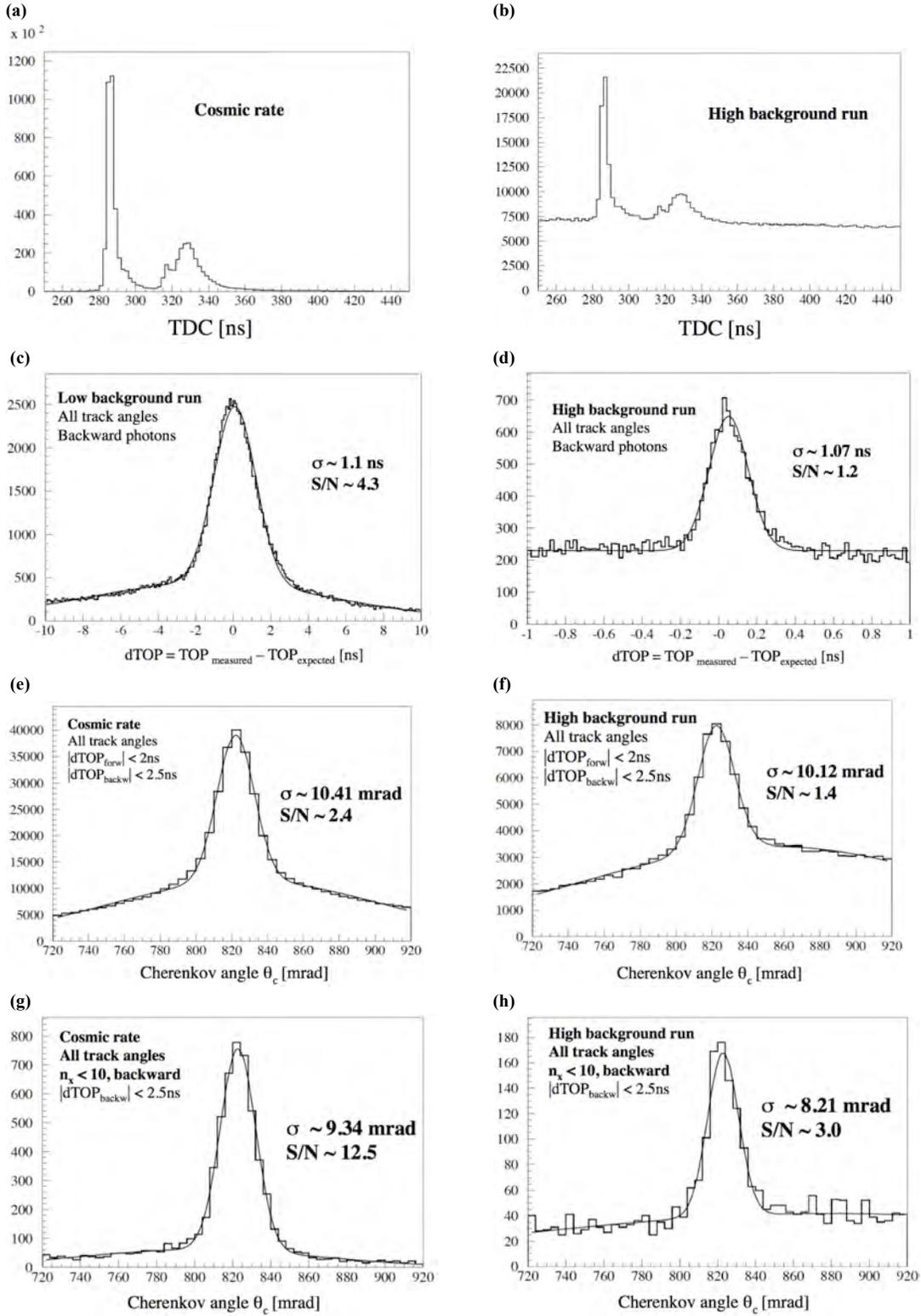

**Fig. 15** (a&b) TDC distribution in "cosmic rate" and "high background" runs. (c&d) dTOP for backward going photons for "cosmic rate" and "high background" runs. (e&f) Cherenkov angle distribution for "cosmic rate" and "high background" runs for all track angles. (g&h) Cherenkov angle distribution in "cosmic rate" and "high background" runs for backward photons only with small number of bounces ($n_x < 10$), which corresponds to the best measured region of the Cherenkov ring in terms of the resolution and the S/N ratio.

## 8. Expected PID performance

Figure 16a shows the MC prediction for the average number of photoelectrons per Cherenkov ring, as a function of the dip angle for a fully-instrumented FDIRC detector at *SuperB*. The plot assumes a nominal design with 48 H-8500 PMTs/bar box.

Figure 16b shows a simple evaluation of the overall PID performance for various detector schemes. Clearly, smaller binning in the y-direction is beneficial to improve the overall PID performance. Hamamatsu encouraged us to use the R11256 tube, which has 3 mm × 12 mm pad sizes, possibly quantum efficiency QE ~36 %, and considerably improved single photon response (pulse height spectrum). The comparison also includes the new Hamamatsu 8 x 8 SiPMT array, where we assumed photon detection efficiency PDE ~52 %; this is shown for comparative purposes only, as this particular photon detector may not be yet ready for the DIRC-like detector application due to its large random noise rate at room temperature and the likelihood of further degradation due to neutron damage.

The present FDIRC prototype provides a single photon Cherenkov angle resolution comparable to the MC simulation. However, converting this result into a measurement of π/K-separation would require additional studies. In particular, any PID algorithm would have to take into account the residual Cherenkov angle distribution tails due to ambiguities, after photon selection and timing cuts have been applied.

This paper is using a cut based data analysis, which is the only way to see projected distributions. However, any real π/K algorithm is likelihood based, which can take advantage of fully correlated probabilities.

(a)
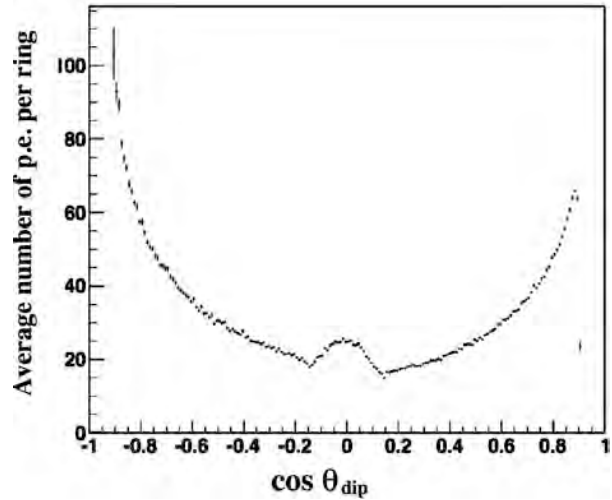

(b)
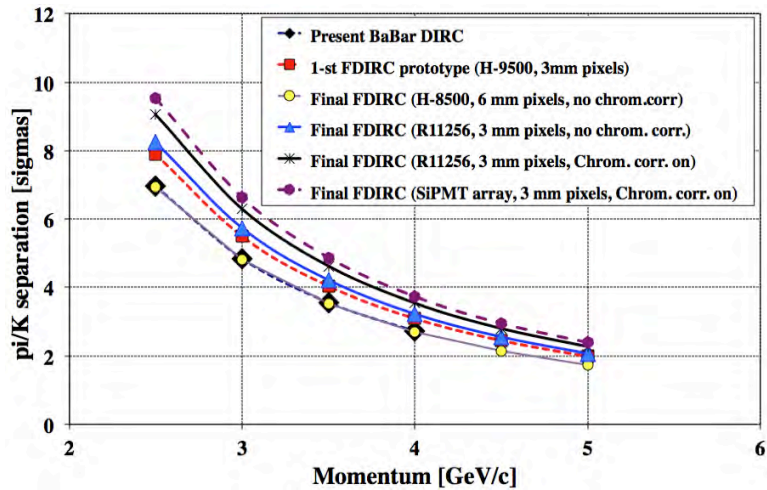

**Fig. 16** (a) MC prediction of the average number of photoelectrons per Cherenkov ring as a function of the dip angle, for a fully instrumented FDIRC detector with H-8500 tubes [1]. (b) Expected π/K performance of the FDIRC detector at *SuperB* for various choices of detectors and pixel sizes [1,14].

## Conclusions

We have successfully constructed and tested a Focusing DIRC prototype detector utilizing a fused silica expansion volume with a focusing element. Using approximately 600,000 cosmic muon triggers at the SLAC CRT, we have been able to demonstrate that the design performs as expected. Using the fast timing of the detector electronics, we have also demonstrated the ability to correct the Cherenkov angle for chromatic dispersion. We have also demonstrated that the FDIRC can reconstruct the Cherenkov angle with good resolution if we add background rate equivalent to ~144 MHz per fully instrumented bar box.

## Acknowledgements

The authors would like to thank M. McCulloch for help in construction, final assembly of the optics and prototype, M. Benettoni for help in the design of the optics enclosure and mechanical support, M. Zago for the mechanical design of the Fbox and the Padova University mechanical workshop personnel for Fbox construction and assembly. We also thank M. Mongelli and V. Valentini of Bari for Fbox mechanical support structure design and M. Franco for helping to construct it, and M. Andrew of the University of Hawaii for electronics board design and support electronics. We thank the France-Stanford Center for Interdisciplinary Studies for its financial support to the R&D project. This work was supported in part by the Department of Energy, Contract DEAC02-76SF00515. We appreciate support of the former *SuperB* collaboration.